\documentclass[conference]{IEEEtran}
\IEEEoverridecommandlockouts
\usepackage{cite}
\usepackage{amsmath,amssymb,amsfonts}
\usepackage{algorithmic}
\usepackage{graphicx}
\usepackage{textcomp}
\usepackage{xcolor}
\usepackage{subfigure} 
\usepackage{hyperref}
\definecolor{b}{rgb}{0, 0, 0}
\definecolor{b1}{rgb}{0, 0, 0}
\definecolor{b2}{rgb}{0, 0, 0}
\usepackage{ragged2e} 
\usepackage{xcolor}
\usepackage{makecell,multirow,diagbox}
\def\BibTeX{{\rm B\kern-.05em{\sc i\kern-.025em b}\kern-.08em
    T\kern-.1667em\lower.7ex\hbox{E}\kern-.125emX}}
\begin{document}

\title{Toward Realization of Low-Altitude Economy Networks: Core Architecture, Integrated Technologies, and Future Directions

\author{Yixian Wang,
        Geng Sun,
        Zemin Sun,
        Jiacheng Wang,
        Jiahui Li,
        Changyuan Zhao,
        Jing Wu,\\
        Shuang Liang,
        Minghao Yin,
        Pengfei Wang,
        Dusit Niyato,~\IEEEmembership{Fellow,~IEEE},\\ 
        Sumei Sun,~\IEEEmembership{Fellow,~IEEE}, and
        Dong In Kim,~\IEEEmembership{Fellow,~IEEE}}
    \IEEEcompsocitemizethanks{\IEEEcompsocthanksitem Yixian Wang, Zemin Sun, Jiahui Li and Jing Wu are with the College of Computer Science and Technology, Jilin University, Changchun 130012, China, and also with the Key Laboratory of Symbolic Computation and Knowledge Engineering of Ministry of Education, Jilin University, Changchun 130012, China (email: yixian23@mails.jlu.edu.cn; sunzemin@jlu.edu.cn; lijiahui@jlu.edu.cn; wujing@jlu.edu.cn).
    \IEEEcompsocthanksitem Geng Sun is with the College of Computer Science and Technology, Jilin University, Changchun 130012, China, and also with the College of Computing and Data Science, Nanyang Technological University, Singapore 639798 (email: sungeng@jlu.edu.cn).
    \IEEEcompsocthanksitem Jiacheng Wang and Dusit Niyato are with the College of Computing and Data Science, Nanyang Technological University, Singapore 639798 (e-mail: jcwang\_cq@foxmail.com; dniyato@ntu.edu.sg).
    \IEEEcompsocthanksitem Changyuan Zhao is with the College of Computing and Data Science, Nanyang Technological University, Singapore, and also with CNRS@CREATE, Singapore 138602 (e-mail: zhao0441@e.ntu.edu.sg).
    \IEEEcompsocthanksitem Shuang Liang and Minghao Yin are with the School of Information Science and Technology, Northeast Normal University, Changchun, 130117, China, (e-mail: liangshuang@nenu.edu.cn; ymh@nenu.edu.cn).
    \IEEEcompsocthanksitem Pengfei Wang is with the School of Computer Science and Technology, Dalian University of Technology, Dalian 116024, China (e-mail: wangpf@dlut.edu.cn).
    \IEEEcompsocthanksitem Sumei Sun is with the Institute for Infocomm Research, Agency for Science, Technology, and Research (A*STAR), Singapore 138632 (e-mail: sunsm@i2r.a-star.edu.sg).
    \IEEEcompsocthanksitem Dong In Kim is with the Department of Electrical and Compute Engineering, Sungkyunkwan University, Suwon 16419, South Korea (e-mail: dongin@skku.edu).}
    \thanks{\textit{Corresponding authors: Geng Sun and Jing Wu.}}
}

\maketitle

\begin{abstract} \textcolor{b}{The rise of the low-altitude economy (LAE) is propelling urban development and emerging industries by integrating advanced technologies to enhance efficiency, safety, and sustainability in low-altitude operations. The widespread adoption of unmanned aerial vehicles (UAVs) and electric vertical takeoff and landing (eVTOL) aircraft plays a crucial role in enabling key applications within LAE, such as urban logistics, emergency rescue, and aerial mobility. However, unlike traditional UAV networks, LAE networks encounter increased airspace management \textcolor{b1}{demands} due to dense flying nodes and potential interference with ground communication systems. In addition, there are \textcolor{b1}{heightened and extended} security risks in real-time operations, particularly the vulnerability of low-altitude aircraft to cyberattacks from ground-based threats. To address these, this paper first explores related} standards and core architecture that support the development of LAE networks. Subsequently, we highlight the integration of technologies such as communication, sensing, computing, positioning, navigation, surveillance, flight control, and airspace management. \textcolor{b}{This synergy of multi-technology} drives the advancement of real-world LAE applications, particularly in improving operational efficiency, optimizing airspace usage, and ensuring safety. Finally, we outline future research directions for LAE networks, such as intelligent and adaptive optimization, security and privacy protection, sustainable energy and power management, \textcolor{b2}{quantum-driven coordination, generative governance, and three-dimensional (3D) airspace coverage}, which collectively underscore the potential of collaborative technologies to advance LAE networks.
%Given the limitations of traditional networks in meeting the multi-layered demands of cross-regional and complex environments, the low-altitude economy (LAE) has ushered in revolutionary progress for urban development and emerging industries through innovative technologies. Moreover, by leveraging unmanned aerial vehicles (UAVs) and electric vertical takeoff and landing (eVTOL) aircraft, it facilitates the efficient and sustainable operation of applications such as urban logistics, emergency rescue, and aerial mobility. In this paper, we first explore the relevant standards and core architecture that support the development of LAE networks. Subsequently, we highlight the integration of technologies such as communication, sensing, intelligent computing, positioning, navigation, surveillance, flight control, and airspace management, thereby demonstrating how this multi-technology integration drives the advancement of real-world LAE applications, particularly in improving operational efficiency, optimizing airspace usage, and ensuring safety. Finally, we outline future research directions for LAE networks, such as intelligent and dynamic optimization, security and privacy protection, and sustainable energy and power, which collectively underscore the potential of collaborative technologies to advance LAE networks.
\end{abstract}


\section{Introduction}
\par With the acceleration of globalization and urbanization, many regions are facing increasingly serious challenges, such as shortages of computational resources and inefficiencies in system performance. For example, the growing complexity of ground transportation systems leads to network congestion, reduced positioning accuracy, and overloaded computational resources, thus affecting the efficiency and responsiveness of urban operations \cite{faheem2024}. At the same time, logistics in suburban and remote areas encounters even more complex issues, such as inconvenient transportation, high costs, and poor timeliness, which severely hamper the flow of goods and emergency response capabilities \cite{Tavasoli2025}. In this context, while traditional unmanned aerial vehicle (UAV)-enabled systems have made progress \cite{Zhao2024, huang2024b, Tong2023a, He2024, Xu2024, Sun2024d, Pan2024}, they still face significant barriers in regulation, technology, and scalability \cite{Mohsan2022}. These systems struggle to effectively meet the multi-layered demands of cross-regional and complex environments, and they are unable to coordinate and provide sustainable services in high-density airspace and resource-constrained areas. Therefore, there is an urgent need for an innovative, green, and intelligent solution that can overcome these \textcolor{b}{limitations. The solution can facilitate} more efficient, flexible, and sustainable technologies to drive long-term development across various sectors.

\begin{figure*}[!hbt] 
	\centering
	\includegraphics[width =7in]{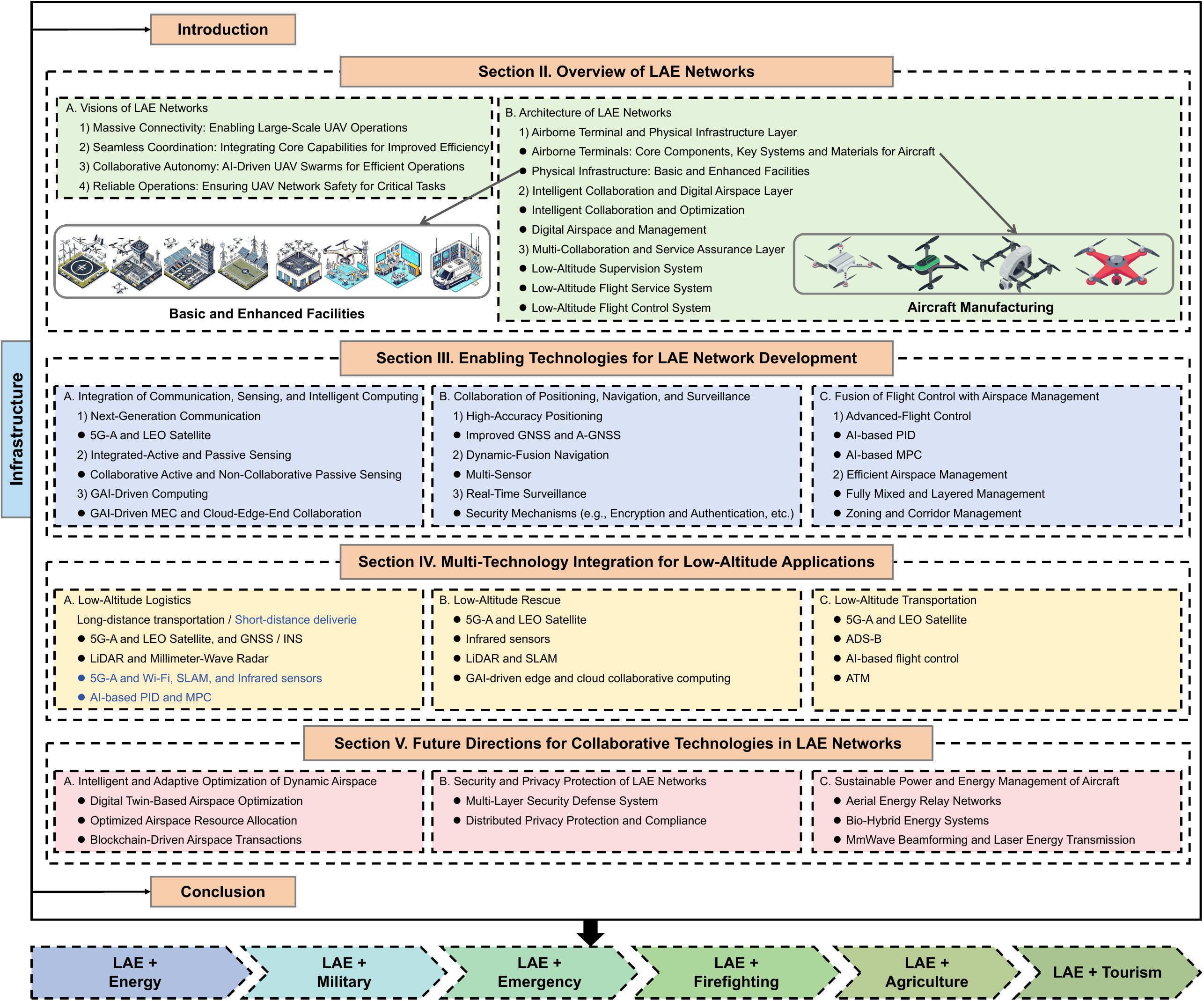}
	\caption{{\textcolor{b}{The survey paper is structured as follows: Overview of LAE Networks (Section II), Enabling Technologies for LAE Network Development (Section III), Multi-Technology Integration for Low-Altitude Applications (Section IV), and Future Directions for Collaborative Technologies in LAE Networks (Section V)}.
    }}
	\label{fig_LAE_Intro_new}
    \vspace{-1.2em}
\end{figure*}

\par As an emerging industrial form, the low-altitude economy (LAE) provides a highly effective solution to current challenges by leveraging their significant potential advantages. Specifically, by unlocking the underutilized strategic airspace between 500 and 3000 meters above the ground and integrating dynamic management with layered operational models, LAE significantly enhances airspace efficiency and flexibility. Moreover, breakthroughs in electric vertical take-off and landing (eVTOL) aircraft \cite{PAVEL2022}, hydrogen-powered propulsion systems \cite{Gao2022}, and distributed thrust technologies \cite{Kim2018} allow these aircraft to offer vertical take-off and landing, low noise levels, and extended range, thereby eliminating the reliance on traditional runways and fuel. In addition, using shared airspace (e.g., the layered operation of logistics aircraft alongside manned aircraft) and reusable infrastructure (e.g., rooftop landing platforms that double as logistics hubs), the LAE ensures efficient use of resources. \textcolor{b}{Furthermore, generative artificial intelligence (GAI) technology has been applied to the optimization of communication and beamforming in LAE networks, which demonstrates the potential of GAI to enhance operational performance and system robustness \cite{zhao2025}}. Given these advantages, LAE demonstrates broad application potential in various sectors. For instance, convenient urban air taxi services can alleviate traffic congestion \cite{Huang2024}. Precise unmanned deliveries can be facilitated in logistics, particularly for e-commerce and medical supplies \cite{Yi2023}. Furthermore, the rapid delivery of goods and personnel in disaster relief can greatly improve emergency response efficiency \cite{zheng2025}. In addition to fostering diverse applications, LAE also drives innovation in UAVs, artificial intelligence (AI), and other technologies. This will further promote the upgrading of high-tech industries and lay the foundation for the digital transformation of the economy \cite{li2024analysis}.

\begin{table*}[!htbp]
    \centering
    \caption{Summary of Related Works vs. Our Survey}
    \label{table0}
    \begin{tabular}{|>{\centering\arraybackslash}m{1.2cm}|m{7.5cm}|m{7.5cm}|}  % \centering\arraybackslash水平居中，m是垂直居中
    \hline
    \textbf{Reference} & \textbf{Key focus of survey} & \textbf{How our survey differs} \\
    \hline
    \cite{Huang2024a} & Designs a low-altitude intelligent transportation system framework that focuses on technologies such as UAVs, AI, and IoT. & Our survey introduces the core architecture of LAE networks and emphasizes how enabling technologies are implemented within such networks.\\
    \hline
    \cite{Jiang2024a} & Explores 3D network coverage and aircraft detection and highlights the role of UAVs in sensing and communication. & We place emphasis on the synergy and integration of technologies, especially in addressing challenges such as aircraft coordination and airspace management, rather than just a single technical issue.\\
    \hline
    %ragged2e 宏包提供了 \justifying，比 \raggedright、\raggedleft 更好用
    \cite{tang2024} & Proposes a collaborative ISAC scheme for low-altitude sensing scenarios to estimate UAV parameters and improve sensing performance. & \multirow{11}{7.5cm}{\justifying Our survey not only focuses on the individual communication technology or the combination of communication and sensing technologies, but also incorporates the computation technology to enhance the overall performance of LAE networks.} \\
    \cline{1-2}
    \cite{ye2025} & Introduces a LAE-oriented ISAC scheme based on deep reinforcement learning that optimizes communication and navigation services. & \\
    \cline{1-2}
    \cite{cheng2024} & Presents an ISAC-based solution that optimizes beamforming at multiple ground stations and UAV trajectory design to improve communication performance and meet sensing requirements. & \\
    \cline{1-2}
    \cite{Li2025a} & Develops a joint communication and interference beamforming design based on a dual-function MIMO cellular system to support authorized user communication and mitigate interference from unauthorized UAVs. & \\
    \hline
    \cite{yang2024} & Investigates the application of EAI in the integration of sensing, communication, computation, and control (ISC3) within LAE. & We further explore the critical roles of navigation, positioning, surveillance, flight control, and airspace management technologies, while examining how these integrated technologies work together to enhance the performance and security of LAE networks. \\
    \hline
    \end{tabular}
\end{table*}

\par However, fully unleashing the potential of LAE and enabling its widespread applications require the development of efficient LAE networks, which must meet several key requirements. 
{\color{b}\begin{itemize}
    \item Intelligent autonomous decision-making capabilities are crucial \cite{yang2024}. This can be achieved through the deep integration of communication, sensing, and computing, which create a real-time environmental sensing and adaptive response system, thereby allowing aircraft to autonomously avoid risks and optimize flight paths in complex airspace. \item A high-precision collaborative technology system is fundamental for efficient operations \cite{Li2025}. This requires the collaboration of multi-modal sensors, anti-jamming navigation, and real-time aircraft monitoring to ensure centimeter-level positioning accuracy, millisecond-level response times, and safe operations in dense airspace. 
    \item An elastic resource scheduling mechanism is key to optimizing operational efficiency \cite{Jiang2024a}. By leveraging dynamic data analysis and digital twin modeling, an intelligent scheduling system can respond to changes in logistics demand or weather conditions, further enhancing airspace utilization and energy efficiency. \item Effective dynamic airspace management is essential \cite{jiang2024}. This involves developing an intelligent air traffic management system that supports the collaborative operation of heterogeneous aircraft. The system should incorporate dynamic airspace allocation algorithms and conflict prediction models for seamless coordination between manned aircraft, logistics UAVs, and ground transportation, while also maintaining compatibility with international airspace management standards.
\end{itemize}
}
\par Based on the above requirements, this survey focuses on the multidimensional synergy of supporting technologies in LAE networks and the key challenges they face, thereby offering comprehensive and in-depth technological insights for the sustainable development of the LAE.

\subsection{Related Works and Contributions}
\label{sec_Related Works and Contributions}
\par Given the popularity of LAE, several surveys have established a solid foundation for understanding the related technologies and applications. Table \ref{table0} summarizes these surveys and shows the distinct contributions of our own survey.

\par In \cite{Huang2024a}, a low-altitude intelligent transportation (LAIT) system framework based on a cyber-physical system (CPS) architecture was proposed, along with a lifecycle management framework to improve the efficiency of low-altitude traffic management. Moreover, the authors also emphasized innovations in key technologies such as UAVs \cite{Liang2024}, AI \cite{Liu2024h}, internet of things (IoT) \cite{Li2024}, 6G \cite{sun2025a}, cloud computing \cite{Sun2025}, and blockchain \cite{Wang2019}, which demonstrate the transformative potential of LAE. Similarly, \cite{Jiang2024a} focused on the core technological components in LAE, with particular attention to 3D network coverage and aircraft detection. Furthermore, the study further elaborated on the role of aircraft in assisting sensing and communication functions, including target detection, broad network coverage, data forwarding, and traffic offloading. Despite these valuable contributions, these studies focus on individual technological aspects and do not delve into how these technologies can be seamlessly integrated and synergized, which is critical for the comprehensive development of LAE.

\par In recent years, research has increasingly focused on the integration of technologies within LAE. For instance, \cite{tang2024} proposed a collaborative integrated sensing and communication (ISAC) scheme for low-altitude sensing scenarios, which aims to cooperatively estimate the parameters of the UAVs and improve the sensing performance. Moreover, \cite{ye2025} introduced a LAE-oriented ISAC scheme based on deep reinforcement learning (DeepLSC) that optimizes communication and navigation services, thus leading to significant improvements in LAE system performance. In a related vein, the authors in \cite{cheng2024} presented an ISAC-based solution for LAE that jointly optimizes beamforming at multiple ground stations and UAV trajectory design to enhance communication performance while meeting sensing requirements. Furthermore, \cite{Li2025a} developed a joint communication and interference (JCJ) beamforming design based on a dual-function MIMO cellular system, which supports communication for authorized users while mitigating the impact of unauthorized UAVs. In addition, the authors investigated how embodied artificial intelligence (EAI) can be applied in the integration of sensing, communication, computation, and control (ISC3) within LAE \cite{yang2024}. Then, the proposed EAI-enabled ISC3 framework was validated through a case study of express delivery services. In conclusion, these studies underscore the diversity and innovation in technological integration within LAE. However, the aforementioned works predominantly emphasize communication, sensing, and computing technologies, while underestimating the essential roles of navigation, positioning, surveillance, flight control, and airspace management.

\par \textcolor{b}{Different} from existing surveys and tutorials, our survey distinguishes itself by focusing specifically on the core technological architecture and applications of LAE networks. Moreover, this survey offers a unique perspective by deeply integrating communication, sensing, intelligent computing, positioning, navigation, flight control, and airspace management technologies. Specifically, it fills a gap in the literature by thoroughly exploring how the synergy of these technologies enables efficient and secure operations in LAE networking, and highlights their critical role in addressing challenges such as aircraft coordination and airspace management in complex airspace scenarios.

\par The main contributions of this paper are summarized as follows:
\begin{itemize}
    \item We highlight the distinctive features of LAE networks compared to traditional UAV networks, and elucidate the core architecture supporting their development. This establishes a robust foundation for the practical implementation and ongoing advancement of LAE.
    \item We explore how the deep integration of technologies such as communication, sensing, intelligent computing, positioning, navigation, surveillance, flight control, and airspace management drives the efficient and safe operation of LAE networks. This provides scalable and self-evolving technical support for the emerging applications such as urban logistics, emergency rescue, and aerial mobility.
    \item We outline future research directions for LAE networks, including intelligent and adaptive optimization, security and privacy protection, sustainable energy and power management, \textcolor{b2}{quantum-driven coordination, generative governance, and 3D airspace coverage}. This underscores the vast potential of collaborative technologies in improving the overall effectiveness of LAE networks from multiple perspectives.
\end{itemize}

\par The structure of this survey is outlined in Fig. \ref{fig_LAE_Intro_new}. Section \ref{sec_Visions and Architecture} presents the vision for LAE networks and its core architecture. Section \ref{sec_Enabling Technologies} provides an overview of the core technologies supporting LAE networks. Section \ref{sec_application} explores the multi-technology integration in real-world applications. Section \ref{sec_Future Directions} discusses the future research directions, and Section \ref{sec_Conclusion} concludes the paper.

\section{Overview of LAE Networks}
\label{sec_Visions and Architecture}
\par In this section, we begin by outlining the visions for LAE networks. Subsequently, we present the core architecture that supports the development and operations of LAE networks. 

\subsection{\textcolor{b}{Visions toward LAE Networks}}
\label{sec_Visions toward LAE Networks}
\par LAE has driven profound transformations in UAV networks, which place higher demands on scalability, efficiency, and reliability. By integrating multifunctional systems and enabling autonomous collaboration, LAE networks achieve highly efficient operations. The following outlines the key aspects of the developmental vision for LAE networks and the standards that support them.

\par 1) {\textit{\textbf{Massive Connectivity: Enabling Large-Scale UAV Operations.}} Traditional UAV networks are typically capable of efficiently supporting only a few dozen UAVs. When this number is exceeded, issues such as signal interference and communication disruptions often arise, which fail to meet the demands of the performance in terms of real-time, flexibility, and reliability \cite{Fotouhi2019}. This is primarily due to the limitations of existing communication protocols and wireless resource management in high-density networks, including frequent channel switching causing interruptions and instability in communication links, and inefficient spectrum management leading to over-utilization or under-utilization of resources \cite{Do2023}. However, LAE networks need to support the coordinated operation of hundreds or even thousands of UAVs, which places greater demands on communication link stability, efficient spectrum utilization, and effective route management.

\par To support such large-scale connectivity, the IEEE 1939.1-2021 standard \cite{1939.1-2021} introduces new technologies that collectively form a structured and scalable low-altitude airspace framework. Specifically, the standard clearly defines communication link quality requirements and network security standards to ensure stable communications in dense environments, and it also optimizes the low-altitude airspace structure and flight route design through communication coverage tests in the target airspace. Moreover, the standard proposes a grid-based route planning method, which divides the airspace into $1\,\text{km} \times 1\,\text{km}$
 grids and plans independent flight paths within each grid to support the coordinated operation of multiple UAVs. Furthermore, the standard integrates enhanced mobile broadband (eMBB) and ultra-reliable low-latency communication (URLLC), thereby guaranteeing reliable performance in high-communication-load scenarios through dynamic frequency adjustment and real-time adaptability.

\par 2) {\textit{\textbf{Seamless Coordination: Integrating Core Capabilities for Improved Efficiency.}} In traditional UAV networks, functions such as sensing, navigation, communication, and control often operate independently rather than being effectively coordinated. For example, the sensing module collects environmental data by using thermal imaging cameras, light detection and ranging (LiDAR), or gas sensors \cite{Fascista2022}, which is then transmitted by the communication module to a ground control center for processing \cite{wang2025a}. Subsequently, the navigation module adjusts the flight path based on the returned results. This multi-step interaction increases the data transmission delays and limits the real-time task performance, especially in complex and dynamic scenarios. In contrast, LAE networks require seamless integration of these functions into a unified framework to enable real-time data exchange and efficient task optimization. 

\par To make this possible, the IEEE 1937.8-2024 standard \cite{1937.8-2024} proposes the seamless integration of these key functions through the coordinated operation of cellular communication terminals and other modules. In particular, the cellular communication terminal, as the core module of the UAV, supports beyond visual line of sight (BVLOS) flight control and real-time transmission of high-definition video data. Moreover, it is also responsible for collecting, processing, and transmitting flight data (e.g., position, speed, and attitude) and the task payload data (e.g., video streams and sensor information). By interacting with ground control centers, the terminal enables the real-time execution of flight commands, as well as precise path planning and navigation. In addition, to ensure efficient data transmission, the standard specifies the use of both serial and Ethernet interfaces, with a requirement that the switching delay of cellular base stations must not exceed 5 milliseconds to maintain continuous communication.

\par 3) {\textit{\textbf{Collaborative Autonomy: AI-Driven UAV Swarms for Efficient Operations.}} Traditional UAV networks heavily rely on human intervention, which requires ground operators to handle task planning, path adjustments, and fault management \cite{Mohsan2022}. This dependency leads to a lack of autonomy in adapting to dynamic environments and unforeseen situations. Moreover, task execution is often decentralized, with UAVs lacking coordination mechanisms and information sharing \cite{Ye2021}, which result in low resource utilization and diminish the effectiveness of task completion. The LAE networks can address the abovementioned issues by incorporating AI into the UAV systems, particularly leveraging artificial general intelligence (AGI) \cite{Goertzel2014} and GAI \cite{Khoramnejad2025} to enable UAVs to autonomously learn, plan, and collaborate \cite{sun2024g}, so that achieving a high level of autonomy within the UAV swarms \cite{Zhang2024b}.

\par To better integrate AGI and enable efficient operations of UAV swarms, the P1954 standard \cite{P1954} defines an architecture and protocol that support self-organizing and spectrum-flexible communication for UAVs, which allows them to automatically establish networks, dynamically adapt spectrum resources, and maintain connections with ground users and devices. The core components include dynamic network deployment, self-organizing spectrum management functions, and integration with existing communication standards such as the IEEE P1900 series. This makes it applicable to disaster emergency communications and the coordination of high-density UAV swarms. In addition, the standard incorporates air-to-air communication specifications from IEEE P1920.1 \cite{1920.1-2022}, along with the dynamic spectrum access and interference co-existence analysis from the IEEE 1900 series, thereby ensuring the reliability and efficiency of communication.

\par 4) {\textit{\textbf{Reliable Operations: Ensuring UAV Network Safety for Critical Tasks.}} Critical tasks such as air taxi services, emergency response, and medical supply transport place high demands on the reliability and safety of UAV networks \cite{Sun2024v, Zheng2024, Li2024a}. However, traditional UAV networks rely on limited communication links for redundant systems, and the adaptability of multi-path technologies is often insufficient due to technical constraints \cite{Li2019}. As a result, UAVs are more prone to communication disruptions or data loss when facing signal interference, network congestion, or other unforeseen events, which can impact task execution and even cause task failure \cite{Chandran2024}. Therefore, to ensure the reliable operation of critical tasks in LAE networks, relevant standards provide comprehensive specifications for flight supervision data transmission and short message transmission protocols.

\par The IEEE 1937.3-2024 standard \cite{1937.3-2024} specifies the content of UAV flight monitoring data, including information such as identity, status, latitude, longitude, altitude, heading, speed, and timestamps. Moreover, it also mandates that the data sampling interval does not exceed 2 seconds to ensure real-time monitoring of UAV flight safety, so that the accurate positioning and behavior tracking can be enabled. Furthermore, the standard utilizes the short message mechanism of the global navigation satellite system (GNSS), which defines the transmission and reception protocols for flight supervision data and specifies the format for the header and data sections of the data packets. In addition, this standard immunity to space, territory, or terrain limitations effectively enhances the continuity and stability of flight supervision data transmission.

\subsection{Architecture of LAE Networks}
\label{sec_Architecture of LAE Networks}
\par To achieve efficient operation of low-altitude aircraft, a well-designed architecture is essential. This architecture supports both system stability and scalability, while facilitating seamless collaboration among its components. Specifically, the details of the overall LAE network architecture are as follows.

%Specifically, Fig. \ref{fig_LAE1} illustrates the overall architecture of the LAE networks, and the details are as follows.

\par 1) {\textit{\textbf{Airborne Terminal and Physical Infrastructure Layer.}} The first layer of the LAE network architecture consists of airborne terminals that are related to the aircraft manufacturing and low-altitude physical infrastructure. Specifically, the former is a crucial component of the aircraft and plays a decisive role in flight performance, safety, and reliability, and the latter focuses on building a robust physical infrastructure to support efficient low-altitude aviation operations.

\par {\textit{\textbf{{Airborne Terminal:}}} The airborne terminal provides the foundation for power support, flight control, environmental sensing, and information exchange in the aircraft, which integrates the core components and key systems that are required for the aircraft operation.
\begin{itemize}
    \item {\textit{Core Components:}} Various components perform unique functions to ensure system stability. Specifically, aviation relays manage the switching of low-power circuits for the safety of the electrical system \cite{Yin2024}. Moreover, engine ignition systems trigger combustion, which affects engine efficiency \cite{Tropina2009}. In addition, circuit protection devices monitor and disconnect faulty circuits to prevent damage, and the aviation contactors control the high-power loads for critical equipment stability \cite{Izquierdo2011}.
    \item {\textit{Key Systems:}} The power system provides strong thrust for flight. For example, small and medium engines are compact and powerful for frequent takeoffs tasks. Moreover, the hybrid systems \cite{Merical2014} offer endurance and quick response for extended flights, and the onboard systems support real-time data exchange and path planning through sensing and navigation. In addition, the flight control systems can manage the attitude and task commands to enhance autonomy and adaptability \cite{Santoso2018}.
\end{itemize}

\par It is worth noting that the key materials for low-altitude aircraft are integral to the development of airborne terminals and aircraft manufacturing \cite{mouritz2012}. In particular, aluminum alloys, known for their high strength-to-weight ratio and strong corrosion resistance, are widely used in aircraft structures, including fixed-wing and helicopters. UAVs are ideal for using carbon fiber due to its high strength and low density. Moreover, composite materials enable the lightweight and optimized the designs for flying cars, such as eVTOLs.

\par {\textit{\textbf{{Physical Infrastructure:}}} A comprehensive and well-rounded operational support system is established to meet the diverse business needs of enterprises. Specifically, such a system includes both basic and enhanced facilities, which not only can support routine operations but also significantly improve the ability of system to respond quickly to unexpected situations and environmental changes.

\begin{itemize}
    \item {\textit{Basic Facilities:}} Takeoff and landing stations are crucial to initiating and completing flight tasks \cite{johnston2020}. Transfer facilities enable smooth transitions between takeoff points and task areas \cite{thipphavong2018}. Moreover, energy stations provide charging or battery replacement services, while emergency landing sites offer the safe options for unforeseen situations \cite{farajijalal2025}.
    \item {\textit{Enhanced Facilities:}} These expand infrastructure are supported by providing secure parking for aircraft, ensuring timely maintenance, and offering support stations for task coordination. Moreover, mobile facilities (e.g., portable charging stations \cite{Ribeiro2022} or repair units \cite{ma2014}) can be quickly deployed to improve the response speed and flexibility.
\end{itemize}

\par 2) {\textit{\textbf{Intelligent Collaboration and Digital Airspace Layer.}} The second layer of the LAE network architecture comprises the intelligent collaboration and optimization, as well as digital airspace and management. The former ensures the real-time flow of information and environmental awareness between low-altitude aircraft and ground facilities, thereby guaranteeing the safety and efficiency of aircraft operations. The latter focuses on the optimal allocation of airspace resources and flight scheduling, which ensures the orderly use of low-altitude airspace and efficient execution of flight operations.

\par {\textit{\textbf{Intelligent Collaboration and Optimization:} }} It is essential for transforming low-altitude airspace into a computable domain, which involves various aspects such as communication, sensing, computing, positioning, navigation, and surveillance. Detailed technical aspects will be discussed further in Section \ref{sec_Enabling Technologies}.

\par {\textit{\textbf{{Digital Airspace and Management:}}} The digital airspace and management provides comprehensive digital services for low-altitude airspace, which covers airspace representation, planning, monitoring, and scheduling, which is detailed as follows:
\begin{itemize}
    \item {\textit{Spatiotemporal Resources:}} Since low-altitude flight tasks operate under varying temporal and spatial conditions, finding the optimal flight paths and time windows within limited resources is crucial \cite{Bouhamed2020}. An spatiotemporal management system can dynamically adjust the requirements of flight tasks and enhance airspace utilization efficiency.
    \item {\textit{City Information Modelling (CIM):}} By integrating geographic information, traffic networks, and building data, a digital model is created to help aircraft understand urban environments \cite{omrany2023}. This enables intelligent path planning and obstacle avoidance, and provides real-time environmental data to enhance flight precision and safety.
    \item {\textit{Knowledge/Rule Base:}} The knowledge base \cite{Doherty2005} compiles information on low-altitude flight, including operational guidelines, airspace regulations, and meteorological data, to provide real-time decision support. The rule base \cite{Stoecker2017} defines standards and regulations to ensure compliant operations and prevent conflicts.
    \item {\textit{Airspace/Flight Management:}} Airspace management plans and monitors low-altitude airspace to prevent collisions or conflicts between aircraft \cite{GUAN2020}. Flight management handles path planning, task execution, and flight status monitoring for aircraft, which ensures timely task completion and compliance with safety regulations \cite{Hamissi2024}.
\end{itemize}

\par 3) {\textit{\textbf{Multi-Collaboration and Service Assurance Layer.}} The third layer of the LAE network architecture meets the regulatory needs of functional departments and the operational requirements of enterprises. By fully integrating digital management and services, this layer empowers governments, air traffic control departments, operators, and businesses to ensure efficient, safe, and compliant low-altitude flights. Moreover, this layer promotes collaboration through information sharing, real-time monitoring, and intelligent analysis to ensure smooth task execution within the LAE networks.

\par {\textit{\textbf{Low-Altitude Supervision System:}} The system links military and civil aviation regulatory agencies with government management platforms, which focus on airspace policy formulation, regulatory enforcement, and compliance oversight. Specifically, the core roles of such systems include real-time monitoring of airspace activity, evaluating the compliance of flight operations, and ensuring airspace safety and order via effective management. Based on this, the system mainly consists of two key modules:
\begin{itemize}
    \item {\textit{Identity Authentication:}} Digital certificates, biometrics, and aircraft codes are used to verify the identity of the aircraft and its crew. Moreover, the system cross-references databases to confirm qualifications and ensure compliance with airworthiness standards \cite{Jiang2020}. Only the certified aircraft and personnel are granted takeoff clearance, thus preventing illegal flights from the outset. 
    \item {\textit{Violation Handling:}} Illegal flights are met with swift legal action, including thorough investigation, evidence collection, and penalties ranging from fines to license suspension or even aircraft confiscation \cite{cracknell2017}. Moreover, violations are made public to deter others and uphold the integrity of low-altitude airspace. 
\end{itemize}
\par {\textit{\textbf{Low-Altitude Flight Service System:}} The system is designed for enterprise users to meet the diverse needs of low-altitude flight operations. Through task planning, route optimization, and resource coordination, it provides customized support services that enhance the efficiency and economic benefits of flight tasks, while effectively reducing operational costs and risks. The system consists of two main modules:
\begin{itemize}
    \item {\textit{Task Scheduling:}} This process integrates aircraft performance, airspace conditions, and weather changes to optimize task plans and flight routes through advanced algorithms. By efficiently scheduling and optimizing the paths, aircraft can complete tasks in the low-altitude environments to minimize the flight time and energy consumption \cite{Banerjee2023, Wang2020}, while improving the efficiency and preventing the resource wastage.
    \item {\textit{Resource Coordination:}} This involves the management and allocation of aircraft, power equipment, and maintenance tools. By using a real-time resource database and intelligent algorithms, dynamic distribution is achieved to ensure timely and efficient supply \cite{Nguyen2019}. Moreover, resource coordination also promotes inter-departmental resource sharing, thereby enhancing resource utilization and emergency response capabilities \cite{DoDuy2021}.
\end{itemize}

\par {\textit{\textbf{Low-Altitude Flight Control System:}} The system connects directly to aircraft and focuses on real-time control and command during flight. The core functions of the system include monitoring flight status, issuing early warnings, and delivering dynamic adjustment commands to ensure safe operations. In emergencies, the system can provide avoidance instructions or initiate emergency return-to-base commands. The system is structured around two essential modules:
\begin{itemize}
    \item {\textit{Flow management:}} This manages the low-altitude airspace to optimize its capacity. By monitoring aircraft numbers, distribution, and flight plans, flow management \cite{balakrishnan2017} can predict traffic changes and implement measures in advance, such as adjusting takeoff intervals or rerouting aircraft to avoid congestion and conflicts.
    \item {\textit{Traffic Control:}} This ensures the flight safety by monitoring low-altitude aircraft and issuing real-time commands. Through communication links, the ground control center sends instructions (e.g., takeoff, landing, altitude changes, and course adjustments) and continuously tracks the flight \cite{spirkovska2017}. Any conflicts or violations are quickly addressed and adjustments are made to maintain smooth operations.
\end{itemize}
{\color{b}
\subsection{Lessons Learned}
\label{sec_Lessons Learned}
\par Building on prior research, LAE networks demonstrate great potential to support large-scale UAV operations, seamless coordination, AI-driven autonomy, and reliable critical tasks. By adopting relevant standards and technical frameworks, LAE networks can effectively tackle the communication and resource challenges that traditional UAV systems face in dense environments. Furthermore, the layered architecture of LAE networks integrates airborne terminals, infrastructure, intelligent collaboration, digital airspace management, and multi-party coordination, which not only provides robust hardware support and real-time data flow, but also guarantees the stability and scalability of low-altitude operations. Despite these strengths, the performance of LAE networks in extreme conditions, such as signal interference or airspace congestion, still requires validation, since robustness and redundancy mechanisms remain inadequate in these scenarios. Therefore, future efforts should focus on enhancing anti-interference capabilities and evaluating performance under uncertainty to meet the diverse demands of LAE.
}
\section{Enabling Technologies for LAE Network Development}
\label{sec_Enabling Technologies}
\par The development of LAE networks is based on three core technologies: (1) integration of communication, sensing, and intelligent computing, (2) collaboration of positioning, navigation, and surveillance, and (3) fusion of flight control with airspace management. This section provides a detailed overview of these technologies.

\subsection{Integration of Communication, Sensing, and Intelligent Computing}
\label{sec_Integration}

\par This integration provides robust technical support for low-altitude aircraft. Communication ensures seamless real-time data transmission, sensing accurately detects the surrounding environment, and intelligent computing leverages big data and AI for rapid analysis and decision making. The synergy of these technologies enables aircraft to quickly adapt to environmental changes, optimize flight paths, and efficiently execute tasks.

\par 1) {\textit{\textbf{Next-Generation Communication.}} As communication demands grow, the conventional 5G networks face challenges such as spectrum constraints, limited coverage, and difficulties in latency control, particularly for the bandwidth, high-reliability, and low-latency applications. To address more complex future needs, \textcolor{b1}{5G-Advanced (5G-A)} \cite{Liberg2024} has emerged as an enhancement and expansion of 5G, which not only inherits key features of 5G, such as eMBB, URLLC, and massive machine type communication (mMTC), but also further integrates new technologies like real-time broadband communication (RTBC), uplink centric broadband communication (uCBC), and harmonized communication and sensing (HCS) \cite{Sufyan2023, Chen2023a, Pang2022}. These technologies can effectively support high-demand applications, including massive IoT \cite{Qi2023, Zhang2023, Xie2022}, autonomous driving \cite{Liu2023c,Liu2019,Lang2025}, and industrial automation \cite{Iradier2021, Bing2024, Tanwar2022}. 

\par In the 5G-A technological architecture, millimeter-wave (mmWave) communication plays a crucial foundational role. As traditional spectrum resources continue to dwindle, the mmWave frequency band (i.e., 30 GHz to 300 GHz) has become a vital technology, which offers large bandwidth and high data rates, thus making it essential for the applications such as ultra-high-definition video streaming and virtual reality \cite{Niu2015, Hong2017}. However, mmWave also faces several challenges, including high propagation loss and poor penetration, which require precise beamforming techniques or dense network layouts \cite{Hamed2018} to compensate for these limitations and ensure stable signal transmission. In this context, massive multiple-input multiple-output (MIMO) \cite{Khan2018, Abdullah2019, Ishteyaq2022} has been widely applied in 5G-A. By utilizing a large number of antenna elements, massive MIMO enables spatial multiplexing and beamforming to enhance signal quality. When combined with mmWave communication, massive MIMO not only mitigates the limitations of mmWave, but also fully leverages its abundant spectrum resources, thereby significantly boosting network capacity and data transmission rates \cite{Rong2018}. 

\par In addition to the traditional techniques, AI and ML approaches have shown great potential in enhancing the performance of mmWave massive MIMO systems. These technologies not only can help optimize the beamforming and channel estimation, but also enable real-time, automated network optimization in dynamic environments. For instance, a deep learning-based hybrid beamforming and channel estimation approach was proposed in \cite{Elbir2022}, which improves spectral efficiency while effectively reducing computational costs. Moreover, \cite{Chen2020c} introduced a neural hybrid beamforming/combining (NHB) scheme that uses autoencoders to optimize beamforming, thereby improving bit error rate performance and system flexibility under the constraints of low-resolution hardware. Furthermore, in \cite{Sudarsan2024}, the authors developed an efficient aided graph neural network combined with hierarchical residual learning (DPrGNN-HrResNetL) to optimize beamspace channel estimation (CE) in mmWave massive MIMO environments. Despite these advancements, challenges such as high hardware complexity and decoding issues caused by low-resolution analog-to-digital converters still remain and should be addressed in future work.

\par Building upon mmWave and massive MIMO technologies, 5G-A also introduces non-orthogonal multiple access (NOMA), which enables more efficient user access through flexible power allocation and multi-user multiplexing strategies on shared frequency resources, further boosting network capacity and stability. For example, the authors divided users into two clusters based on channel correlation and employed a hybrid analog-digital beamforming scheme with power allocation (PA) optimization, thus effectively improving system energy efficiency \cite{Zeng2019}. Another approach involved the intelligent reflecting surface (IRS)-assisted mmWave beamspace NOMA, where joint optimization of both active and passive beamforming enhanced the weighted sum rate \cite{Liu2021c}. In addition, the authors proposed an angular domain NOMA scheme based on user scheduling and precoder design \cite{Shao2020}, which optimizes the precoder and decoder to improve overall system throughput and quality of service (QoS). As such, the integration of these techniques enhances network performance and efficiency, while simultaneously providing robust technological support for LAE networks.

\par Additionally, due to the relatively low orbital altitude (i.e., ranging from 500 to 2000 kilometers), low Earth orbit (LEO) satellites experience lower latency and path loss \cite{Su2019, Qu2017, Lin2021}, which leads to more stable communication signals, which is especially suitable for the needs of low-altitude aircraft over wide areas. To fully leverage the potential of LEO satellites, several innovative technical solutions have been proposed. For instance, the LEO satellite selection-based computation offloading (LSSBCO) algorithm optimized the satellite selection to reduce computation offloading delay and improve the communication efficiency \cite{Zhang2024a}. In terms of channel allocation, an improved artificial bee colony (IABC) algorithm was introduced to mitigate beam interference in LEO satellite networks, optimize channel utilization, and reduce propagation delay, thereby enhancing the system throughput \cite{Zhou2023}. Furthermore, the adaptive coverage enhancement (ACE) approach improved the success rate of random access and reduced power consumption, significantly boosting the performance of Narrowband IoT (NB-IoT) in LEO satellite networks \cite{Hong2024}. This approach also reduces the performance gap by approximately 58\% compared to the traditional 3GPP model. In summary, LEO satellite technology enhances the communication quality, optimizes the resources, and plays a key role in computation offloading and channel allocation, which makes it essential for the LAE networks. However, as the numbers of aircraft and the task complexity grow, frequent switching of LEO satellites may complicate system management and require further optimization.

\par 2) {\textit{\textbf{Integrated-Active and Passive Sensing.}} Sensing technologies can gradually evolve towards the integration of collaborative active sensing and non-collaborative passive sensing \cite{varotto2021}. On the one hand, collaborative active sensing relies on the cooperation of multi-modal sensing technologies, which integrate various sensors, such as radar, LiDAR, vision, and infrared to comprehensively perceive the environment from multiple dimensions. The complementarity of these sensors overcomes the limitations of individual sensors \cite{Liu2023, Wang2025}, which significantly improves the sensing accuracy and enhancing target detection, recognition, and tracking precision. For example, the authors designed a feature registration and complementary sensing network (FRCPNet) \cite{Ji2025}, which aligns multi-level features at the pixel level and enhances the semantic correlation between modalities, thereby improving the accuracy of salient object detection in combined RGB and thermal images. Moreover, in \cite{Clar2025}, a visual transformer-based model (ViTAL-TAPE) that combines photometric and radiometric data was introduced to overcome the limitations under varying lighting and viewing conditions, ultimately achieving high-precision landing detection with an accuracy of 0.01 meters.

\par On the other hand, collaborative active sensing thrives on the seamless integration of sensing and communication systems to enhance both the accuracy and timeliness of environmental perception. In this scheme, active sensing technologies are used to collect real-time environmental data, while the communication system efficiently transmits and shares this data, thereby improving the responsiveness and information flow of the sensing system. For instance, in \cite{Chen2020}, the authors proposed a joint sensing-communication (JSC) collaborative sensing UAV network (CSUN), which synchronizes radar sensing and data communication through beam sharing, thus strengthening the collaborative sensing capacity. In another study \cite{Mohammadkarimi2024}, the authors combined secondary surveillance radar (SSR) with transponder-equipped air obstacles to enable the safe navigation of small UAVs in the airspace by sharing altitude and identification code information. Furthermore, ISAC is gaining momentum as a cutting-edge research direction, aiming to achieve efficient resource sharing and boost system performance. In one example, the authors presented a multi-UAV-supported IoT system that employs ISAC services and integrates radar mutual information for assessing sensing performance, which significantly enhances communication rates and ensures fairness in communication \cite{Liu2024l}. In addition, the authors shared a joint approach for UAV maneuvering and transmission beamforming, so that maximizing the communication throughput while meeting the necessary sensing beam gain requirements \cite{Lyu2023}. Despite the progress made, balancing the distinct needs of sensing and communication systems in high mobility, multiple targets, and environmental clutter, while ensuring effective target detection and precise communication performance, remains a key challenge that requires further optimization.    

\par Non-collaborative passive sensing involves utilizing existing wireless communication signals (e.g., Wi-Fi or 4G/5G signals) to detect the surrounding environment. Unlike active radar, passive sensing captures reflected waves from signals emitted by wireless communication systems to gather environmental information, without the need for dedicated transmission signals \cite{Lima2021}. The main advantage of this technology is that it does not interfere with existing communication systems and requires no additional hardware, which makes it ideal for privacy-preserving and cost-effective monitoring applications \cite{Savazzi2019}. However, weak reflected signals in passive sensing are susceptible to multi-path effects, static clutter, and noise. Additionally, extracting useful information from these complex signals and accurately detecting moving targets without active control remains a challenge.

\par Consequently, the integration of active and passive sensing will become the core of intelligent sensing systems. By complementing each other, the system can leverage active radar for high-precision distance and velocity data, while also using the environmental changes detected by passive sensing to form a richer and more accurate perception. For example, in LAE applications, active radar measures real-time distance between UAVs and other aircraft, thus providing precise flight trajectory information, while passive sensing aids in detecting and tracking surrounding ground targets, such as vehicles or buildings. This collaboration not only improves the perception accuracy and robustness, but also optimizes the resource utilization. However, fully exploiting the benefits of this collaboration requires overcoming challenges such as signal interference, real-time system coordination, and multi-modal data fusion.

\par 3) {\textit{\textbf{GAI-Driven Computing.}} Traditional AI plays a crucial role in mobile edge computing (MEC) and cloud-edge-end collaboration. Specifically, by incorporating deep reinforcement learning (DRL) or graph neural network (GNN), MEC can optimize computation offloading and resource allocation, thereby enhancing task processing efficiency while reducing latency and energy consumption \cite{Liu2020, Zhao2023}. Moreover, the cloud-edge-end collaborative framework boosts system flexibility and responsiveness \cite{Liu2024f, Liu2024g}. However, these AI faces several challenges in practical applications. First, UAV tasks often take place in highly uncertain environments, in which traditional AI reliant on historical data struggles to handle unforeseen events or unknown scenarios. Second, traditional AI increases computational latency when processing large amounts of data, which impacts real-time decision-making and makes resource scheduling more complex as the task load grows. Finally, the high-dimensional data collected by multiple sensors often causes traditional AI to lose key information during feature extraction, thereby affecting decision quality. As a solution, GAI \cite{sun2025h, he2024a, sun2024n} emerges as a key technology to address these issues, and it is expected to play a significant role in the LAE networks.

\begin{figure}[!t]
    \centering
   \includegraphics[width=3.5in]{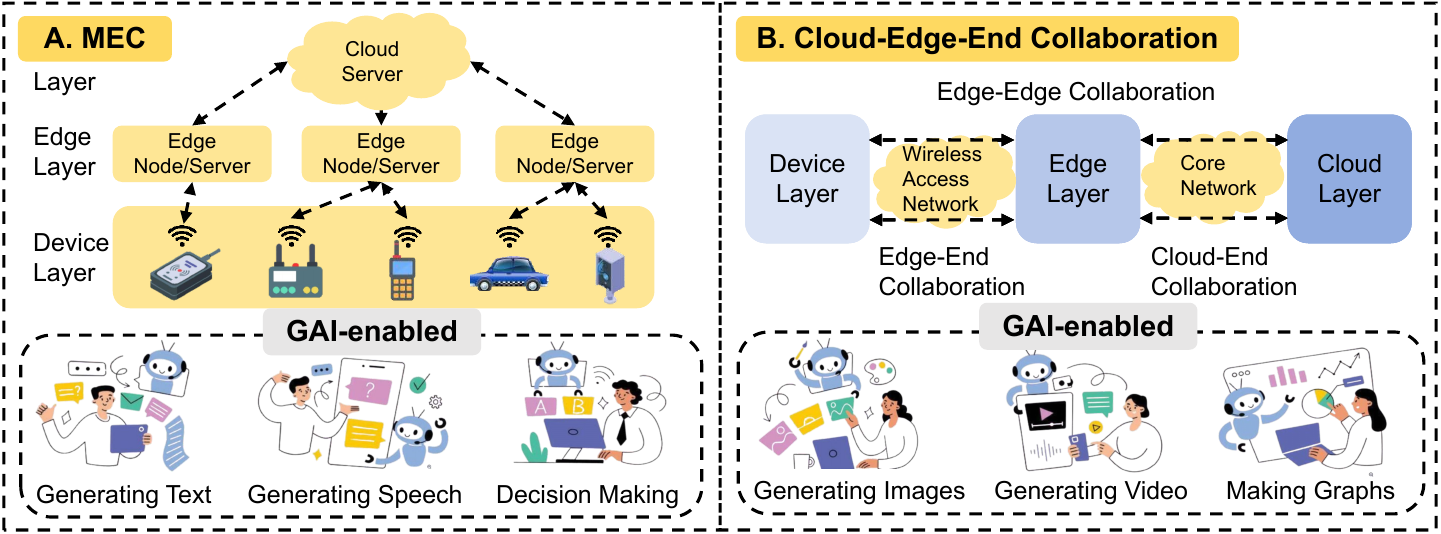}
    \caption{\textcolor{b}{GAI-driven computing in LAE networks. \textit{Part A} presents an MEC-based architecture, where GAI supports applications including text generation, AI chatbots, and decision-making. \textit{Part B} illustrates a cloud-edge-end collaborative framework, where GAI facilitates functions such as image generation, video generation, and graph creation.}}
    \label{fig_LAE_MEC_Cloud}
    \vspace{-3.5ex}
\end{figure}

\par In MEC systems, GAI leverages its powerful modeling and predictive capabilities to tackle the optimization of task offloading, resource allocation, and other aspects in dynamic environments, \textcolor{b}{as shown in Fig. \ref{fig_LAE_MEC_Cloud} \textit{Part A}}, thereby improving the computational efficiency and overall system performance. For example, a robust multi-access edge computing framework for UAV and ground station was designed, which employs the GAI-enhanced heterogeneous agent proximal policy optimization algorithm to improve the accuracy of the actor network modeling in complex environments and effectively mitigates instability issues in multi-agent reinforcement learning \cite{you2025}. Moreover, the authors proposed a new MEC offloading decision and resource allocation scheme that combines generative diffusion model (GDM) and DRL, thus reducing the model training costs and achieving significant optimization performance in terms of the task processing delay and energy consumption \cite{du2024}. For the mobile augmented reality (MAR) applications in MEC, a GAI-based super-resolution control scheme was introduced to optimize the energy efficiency and enhance the service satisfaction \cite{Na2024}. Furthermore, in \cite{Liu2024m}, the authors developed an MEC-enabled GDM-based artificial intelligence-generated content network, which integrates the joint user association and computational offloading game model to effectively address the communication and computational scheduling issues.

\begin{table*}[!htbp]
    \centering
    \caption{Summary of the Integration of Communication, Sensing, and Intelligent Computing Technology}
    \label{table1}
    \begin{tabular}{c|c|c|c}
    \hline
    \textbf{} & \textbf{Technologies} & \textcolor{b}{\textbf{References}} & \textbf{Pros \& Cons} \\
    \hline
    \multirow{6}{*}{\shortstack{\textbf{Next-Generation} \\ \textbf{Communication}}}
    & \multirow{3}{*}{5G-A} & \multirow{3}{*}{\cite{Elbir2022, Chen2020c, Sudarsan2024, Zeng2019, Liu2021c, Shao2020}} & \checkmark Increasing network capacity and data transmission rates \\
    & & & \checkmark Achieving automated network optimization in dynamic environments\\
    & & & \texttimes Requiring extensive resources for mmWave beamforming and network layout \\
    \cline{2-4}
    & \multirow{3}{*}{LEO satellite} & \multirow{3}{*}{\cite{Zhang2024a, Zhou2023, Hong2024}} & \checkmark Providing stable communication signals for wide-area coverage  \\
    & & & \checkmark Improving efficiency, optimizing resources, and boosting throughput \\
    & & & \texttimes Frequent switching of LEO satellites potentially complicating system management\\
    \hline
    \multirow{6}{*}{\shortstack{\textbf{Integrated-Active} \\ \textbf{and}\\ \textbf{Passive Sensing}}}
    & \multirow{3}{*}{\shortstack{Collaborative \\ Active Sensing}} & \multirow{3}{*}{\cite{Ji2025, Clar2025, Chen2020, Mohammadkarimi2024, Liu2024l, Lyu2023}} & \checkmark Overcoming single-sensor limitations, improving detection accuracy \\
    & & & \checkmark Enhancing sensing accuracy and timeliness\\
    & & & \texttimes Balancing sensing and communication demands is challenging \\
    \cline{2-4}
    & \multirow{3}{*}{\shortstack{Non-Collaborative \\ Passive Sensing}} & \multirow{3}{*}{\cite{Lima2021, Savazzi2019}} & \checkmark Avoiding interference with communication systems, and reducing hardware needs \\
    & & & \checkmark Enabling privacy-preserving and cost-effective monitoring applications \\
    & & & \texttimes Struggling with weak reflected signals, multi-path effects, and noise \\
    \hline
    \multirow{7}{*}{\shortstack{\textbf{GAI-Driven}\\ \textbf{Computing}}}
    & \multirow{3}{*}{GAI-Driven MEC} & \multirow{3}{*}{\cite{you2025, du2024, Na2024, Liu2024m}} & \checkmark Significantly optimizing task processing delay and energy consumption \\
    & & & \checkmark Enhancing service satisfaction while reducing model training costs\\
    & & & \texttimes Having limited computational resources while failing to meet strict real-time requirements \\
    \cline{2-4}
    & \multirow{3}{*}{\shortstack{GAI-Driven \\Cloud-Edge-End\\ Collaboration}} & \multirow{3}{*}{\cite{Zhou2024, Hu2024}} & \checkmark Enhancing system stability and robustness through strong adaptability\\
    & & & \checkmark Improving overall system efficiency and response speed by cross-layer optimization \\
    & & & \texttimes Increasing complexity and operational costs due to higher implementation and maintenance\\
    \hline
    \end{tabular}
\end{table*}

\par To further enhance the adaptability and responsiveness of the system, the cloud-edge-end collaboration architecture has emerged to provide a more flexible solution, \textcolor{b}{as shown in Fig. \ref{fig_LAE_MEC_Cloud} \textit{Part B}}. In this architecture, GAI optimizes resource scheduling and task offloading across cloud and edge nodes through intelligent prediction and dynamic adjustments, thereby ensuring efficient system coordination and low-latency performance in complex scenarios. For instance, in \cite{Zhou2024}, the authors presented a novel edge-cloud deployment scheme that optimizes service delay of large language models (LLMs) through radio resource allocation and task offloading, while improving the generation service quality by using an in-context learning method. In addition, a cloud-edge collaborative advanced driver assistance (ADAS) system based on multi-modal LLMs was introduced to enhance the understanding of the traffic scene and the decision-making capabilities of CogVLM2 and ChatGPT-4o by integrating LoRA fine-tuning and few-shot learning techniques \cite{Hu2024}. Based on this, latency, energy consumption, and QoS were significantly optimized. Although GAI-driven collaboration architecture enhances flexibility and efficiency, its high implementation and maintenance costs, especially with frequent task scheduling and resource allocation, may increase the system complexity and operational expenses.

\par In conclusion, Table \ref{table1} illustrates the applicability of communication, sensing, and intelligent computing in LAE. By adopting these advanced technologies, efficient data transmission and stable connectivity, comprehensive environmental sensing, and intelligently assisted decision-making can be achieved. However, relying solely on individual technologies or limited combinations of them is no longer sufficient to meet the growing demand. 
%Therefore, integration of communication, sensing, and intelligent computing has become essential to enhance system efficiency and increase intelligence levels. 

\par Numerous studies have explored the integration of these three technologies. For example, \cite{Qi2022} proposed a unified integrated sensing, communication, and computation (ISCC) framework, aiming to optimize the limited system resources in 6G wireless networks. In \cite{Xu2023a}, the authors investigated a UAV-enabled ISCC (U-ISCC) framework and analyzed the trade-off between the computational capability and beam-sensing gain. Moreover, a novel wireless scheduling architecture was introduced \cite{Zhao2022} (Fig. \ref{fig_LAE_ISCC}) that jointly optimizes the coordinated gains of sensing, communication, and computing to address heterogeneous demands in ISCC. In the context of IoT and cellular networks, \cite{Qi2021} examined the ISCC challenges in beyond-5G cellular IoT through two joint beamforming design algorithms. Furthermore, in \cite{Li2023b}, the authors presented AirComp as a communication approach and designed two MIMO ISCC over-the-air (ISCCO) schemes (i.e., separated and shared) to enhance performance. In edge computing, \cite{Wen2024} studied a multi-device edge AI system that integrates AI model splitting with ISAC to deliver low-latency intelligent services. In addition, \cite{Wang2023} developed a NOMA-assisted joint communication, sensing, and multi-tier computing framework. Despite these developments, several challenges persist in real-world LAE applications. Specifically, the rapid growth of the IoT devices and the limited communication resources complicate the access and transmission of large-scale devices. {\color{b2} Moreover, the air interface between UAVs and ground control systems in LAE networks is inherently dynamic and open, which increases the complexity of achieving resilient coverage and secure communication in a three-dimensional (3D) space. As a result, the seamless integration of aerial vertical coverage with terrestrial cellular networks has become a pressing issue.
%Moreover, balancing the computational load, energy efficiency, and task performance remains critical, while designing robust optimization algorithms for dynamic, complex environments continues to be a major research challenge.

\par To address this, the careful planning of spectrum resources and the introduction of dynamic interference management mechanisms are essential to improve spectrum efficiency, particularly in the context of co-existence with existing 5G infrastructure. For example, the study \cite{Yang2024a} demonstrated that reconfigurable intelligent surface (RIS)-assisted joint beamforming methods can intelligently control spatial degrees of freedom under limited power constraints, which effectively suppresses co-channel interference in spectrum reuse scenarios and reduces sensing errors by more than 46.7\%. Moreover, the simultaneously transmitting and reflecting surface (STARS)-assisted orthogonal frequency division multiple access (OFDMA) ISCC network was introduced to jointly optimize subcarrier allocation and reflection/transmission beamforming, so that sensing, communication, and computing functions share spectrum efficiently in wide-bandwidth scenarios \cite{Li2024f}. In addition, the authors proposed a spectrum prediction method by using the multi-scale feature fusion graph convolutional network (MFGCN) to improve the management of spectrum and the efficiency of ISCC systems \cite{Li2024g}. Inspired by these advances, future LAE network deployment can leverage ISCC capabilities to construct a dynamic radio environment map that captures spectrum states and interference patterns. This approach will enable joint optimization of multiple access and interference management, ultimately facilitating intelligent spectrum sharing and resource scheduling in air-ground integrated environments.
}

\begin{figure}[!t]
    \centering
   \includegraphics[width=3.5in]{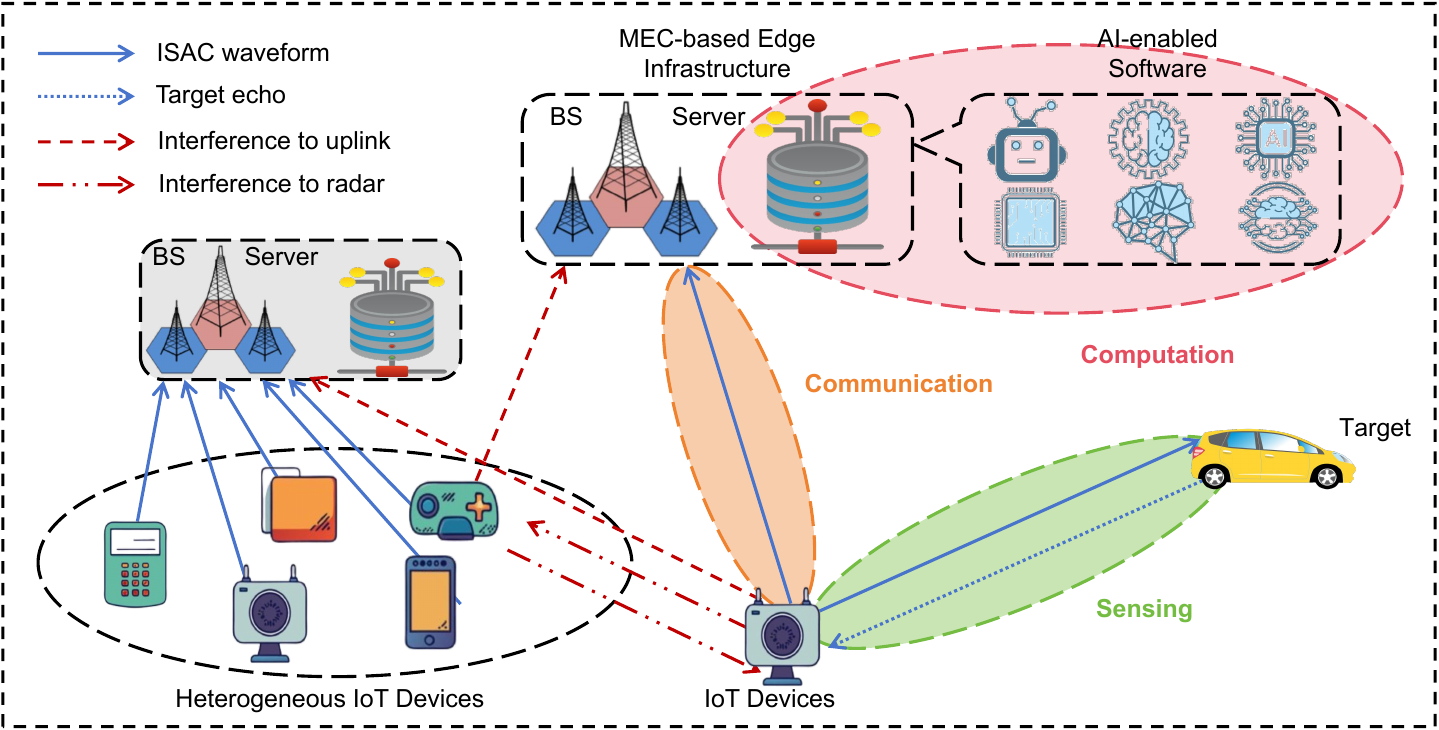}
    \caption{Proposed multi-user wireless network architecture under the ISCC paradigm in \cite{Zhao2022}. Specifically, the BS utilizes ISAC waveforms to integrate communication and sensing while collaborating with IoT devices for computational tasks. Moreover, MEC-based edge infrastructure and AI-enabled software provide intelligent decision support.}
    \label{fig_LAE_ISCC}
    \vspace{-3.5ex}
\end{figure}

\subsection{Collaboration of Positioning, Navigation, and Surveillance}
\label{sec_Collaboration}

\par This collaboration forms the foundation for flight stability and precise operations in the LAE. Positioning ensures accurate location tracking in complex environments, navigation provides optimal route planning, and surveillance continuously monitors the airspace, thus identifying potential risks in real time. The coordination of these technologies greatly increases the situational awareness of the aircraft and reinforces the safety defenses of the system.

\par 1) {\textit{\textbf{High-Accuracy Positioning.}} The GNSS is the cornerstone of modern positioning technology. Specifically, GNSS determines accurate 3D positioning (i.e., longitude, latitude, and altitude) by receiving radio signals from multiple satellites and calculating the distance between the receiver and each satellite, along with precise time synchronization. Common GNSS systems include GPS, GLONASS, Galileo, and BeiDou. They have played a crucial role in a wide range of applications, such as intelligent transportation systems (ITS) with connected autonomous vehicles (CAVs) \cite{Adegoke2019, Elghamrawy2022, Rahman2022}, as well as in complex environments such as urban canyons \cite{Ng2021, More2024, Weng2024}.

\begin{figure}[!t]
    \centering
   \includegraphics[width=3.5in]{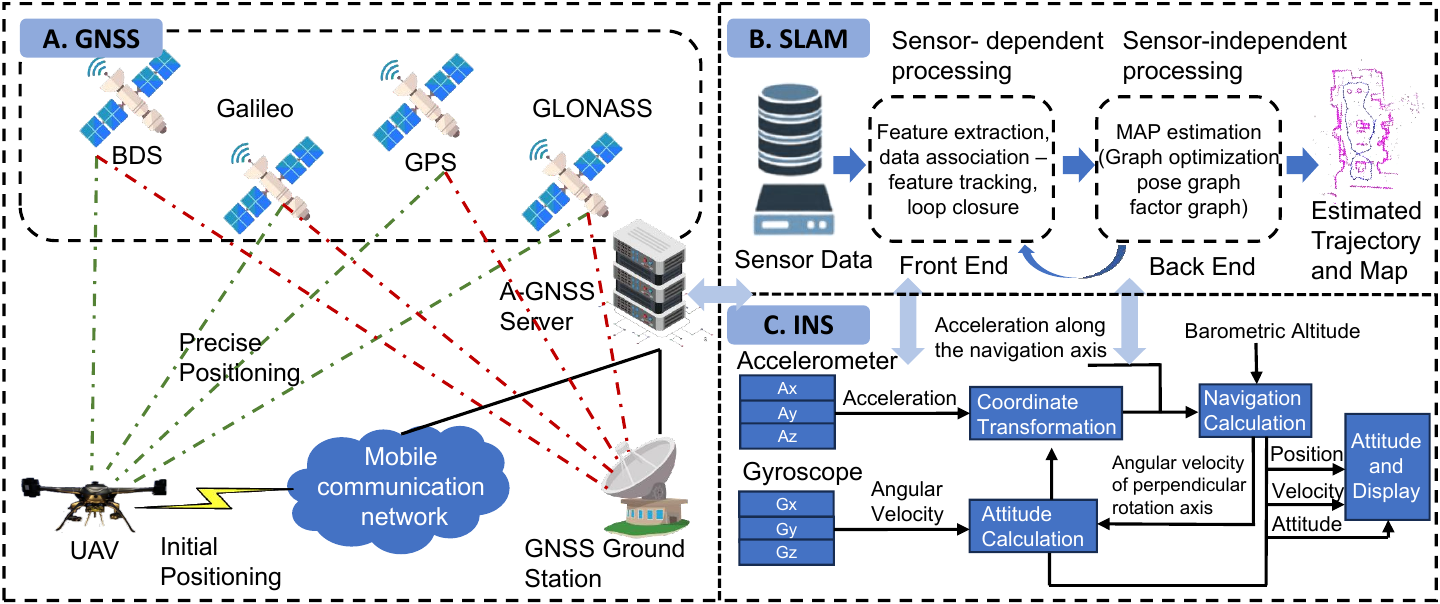}
    \caption{\textcolor{b}{An integrated UAV navigation and positioning framework. \textit{Part A} presents the GNSS module, which utilizes BDS, Galileo, GPS, and GLONASS for initial positioning, with A-GNSS and ground stations enhancing accuracy. \textit{Part B} outlines the SLAM process, where the front end handles feature extraction and data association, while the back end performs map estimation and pose optimization. \textit{Part C} depicts the INS module, which leverages accelerometer and gyroscope data for attitude and navigation calculations to determine position, velocity, and orientation.}}
    \label{fig_LAE_GNSS_INS_SLAM}
    \vspace{-3.5ex}
\end{figure}

\par To further enhance the performance of GNSS, researchers have explored innovative technologies in various fields, such that significantly increasing the accuracy and reliability of the GNSS positioning. In the agricultural vehicle sector, \cite{Xie2024} proposed a collaborative positioning algorithm that integrates vehicle motion state data and GNSS observations by using an extended Kalman filter (EKF), thus boosting the performance of standalone GNSS positioning. Subsequently, based on Bayesian theory, the author combined GNSS data, ultra-wideband (UWB) ranging data, and inter-vehicle relative positioning information to further enhance the positioning accuracy. Moreover, the authors in \cite{Tahir2019} developed a generalized theoretical framework for analyzing and measuring vehicle distances in workshops, which enables precise distance measurement by exchanging GNSS observation values. For complex urban environments, challenges such as signal blockage, multi-path effects, and non-line-of-sight (NLOS) signals may arise. To address these issues, \cite{Yao2022} introduced a vision-aided approach that enhances GNSS positioning by identifying and eliminating NLOS satellite signals in urban environments. Furthermore, in \cite{Liu2024}, the authors presented a joint positioning algorithm that combines Doppler measurements from LEO satellites with GNSS data, so that improving the positioning accuracy to 250 meters. To further support the research, a high-fidelity GNSS urban simulation tool was designed, which uses ray-tracing techniques to replicate signal interference in urban environments \cite{Zhang2021}, thereby helping to precisely analyze signal characteristics and optimize positioning systems. 

\par The successful application of innovative technologies in the aforementioned fields provides robust technical support for the positioning needs in LAE. Additionally, the introduction of assisted GNSS (A-GNSS) technology \cite{van2020, AbuShaban2020, Li2023} has become a crucial solution to further boost the positioning performance, \textcolor{b}{as shown in Fig. \ref{fig_LAE_GNSS_INS_SLAM} \textit{Part A}}. Specifically, A-GNSS enhances the overall performance of GNSS by utilizing terrestrial mobile communication networks to transmit real-time satellite ephemeris, frequency corrections, and clock calibration data. This allows GNSS receivers to acquire and track satellite signals more quickly, thereby accelerating the signal search process and shortening the time to first fix (TTFF). Furthermore, A-GNSS offers corrections for ionospheric delays, satellite clock errors, and orbital inaccuracies, which minimize the signal propagation errors and significantly improve positioning accuracy. As a result, both the improved GNSS and A-GNSS will play a critical role in meeting the high-precision positioning requirements of low-altitude operations within the LAE networks.

\par 2) {\textit{\textbf{Dynamic-Fusion Navigation.}} SLAM and inertial navigation system (INS) play a crucial role in complex environments, where GNSS signals are unstable or unavailable. SLAM can equip camera \cite{Song2022} or LiDAR \cite{He_2024} sensors to construct environmental maps in real-time and perform self-localization, while integrating GNSS data to improve positioning accuracy and robustness, \textcolor{b}{as illustrated in Fig. \ref{fig_LAE_GNSS_INS_SLAM} \textit{Part B}}. Moreover, INS uses the accelerometer and gyroscope in the inertial measurement unit (IMU) to measure the acceleration and angular velocity of the aircraft in real time, \textcolor{b}{as shown in Fig. \ref{fig_LAE_GNSS_INS_SLAM} \textit{Part C}}. This internal measurement approach enables the INS to navigate reliably even in the absence of external signals \cite{YU2024, Zhang2019a}. However, single-sensor systems often have limitations. Specifically, cameras may struggle in low-light or high-speed conditions, which leads to inaccurate feature detection and difficulties tracking moving objects. Although LiDAR provides precise depth information, it may encounter limitations when dealing with reflective or transparent surfaces, such as windows or water, thereby causing it to produce erroneous measurements or fail to detect objects properly. In addition, INS may experience drift in the absence of external reference points, such as when GNSS signals are weak, unavailable, or obstructed, thus leading to cumulative positioning errors that worsen over time. Therefore, by combining sensor data, the navigation system leverages their strengths to provide more accurate and robust navigation than single-sensor systems.

\par Several studies have highlighted the benefits of multi-sensor fusion in improving navigation performance. For example, the integration of semantic information, laser point clouds, and GNSS data into the visual SLAM framework (SLG-SLAM) \cite{Wu2024} was proposed to address the limitations of existing optimization algorithms in complex driving scenarios. Through experiments on four public datasets (i.e., KITTI \cite{Geiger2012}, multi vehicle event (MVE) \cite{Zhu2018}, M2DGR \cite{Yin2022}, and complex urban (CU) \cite{Jeong2018}) and one self-collected dataset \cite{Liu2024c}, SLG-SLAM significantly improves the localization accuracy, with the average absolute trajectory error reduced by 43.50\% and 14.91\% on the KITTI and MVE datasets, respectively. Moreover, in \cite{s22124327}, the author developed an EKF-based INS/LiDAR SLAM navigation system through a robust loose coupling approach, which achieves better localization accuracy than pure INS navigation (Fig. \ref{fig_LAE_GNSS_INS}). Furthermore, the authors designed a tightly coupled approach that integrates visual, IMU, and GNSS data to jointly optimize image reprojection, IMU pre-integration, and GNSS measurement errors \cite{Liu2021}. Notably, the study highlights that this approach outperforms state-of-the-art visual-inertial SLAM, GNSS single-point positioning, and loose coupling approaches, particularly in environments dominated by low-rise buildings and urban canyon scenarios. Consequently, innovative navigation systems based on multi-sensor fusion are key technologies for achieving high precision, reliability, and autonomy in the future of the LAE. However, this fusion may face challenges such as higher computational complexity, sensor noise that affects accuracy, and data synchronization issues, which require further attention in the context of the LAE networks.

\begin{figure}[!t]
    \centering
   \includegraphics[width=3.5in]{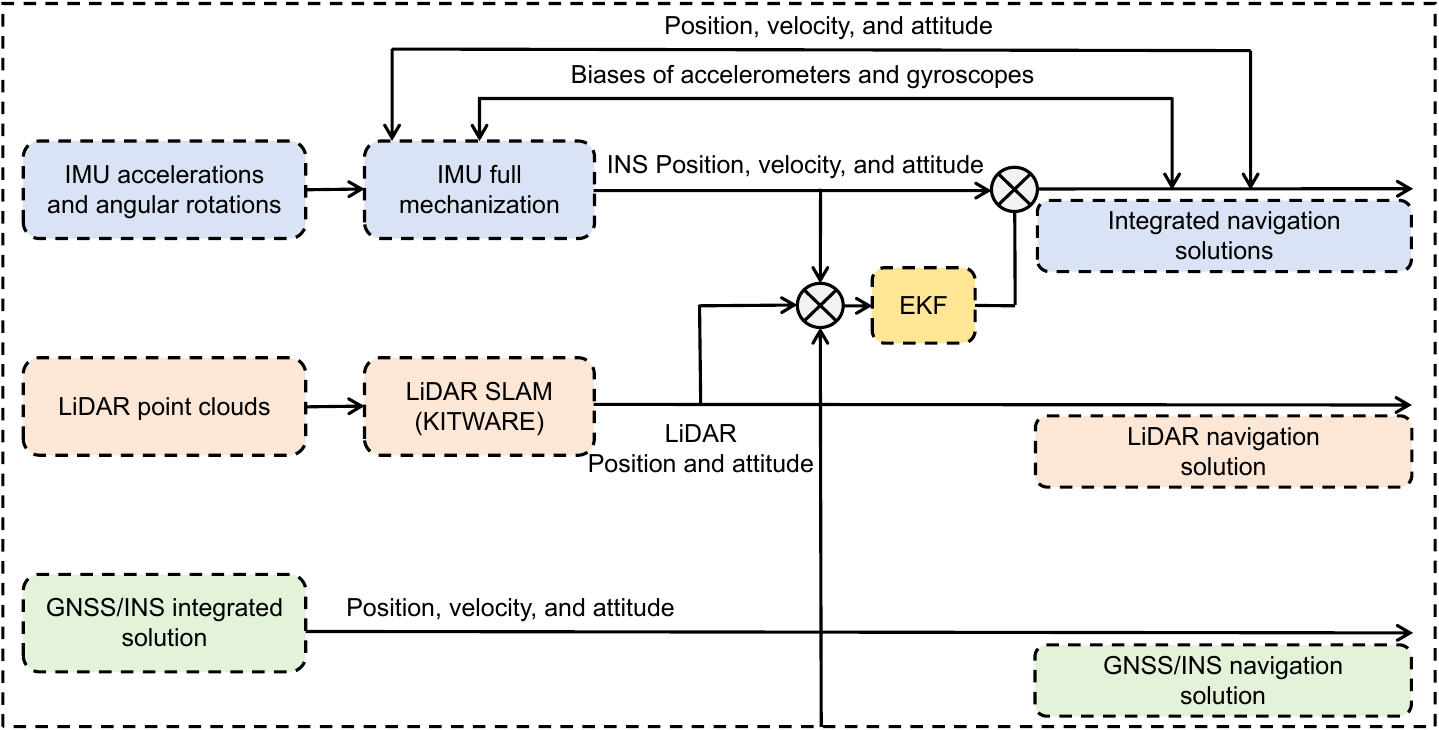}
    \caption{Overall structure of the INS/LiDAR SLAM LC integration system in \cite{s22124327}. Specifically, the IMU module processes acceleration and angular velocity data to compute the inertial navigation solution. Moreover, the LiDAR SLAM module extracts position and attitude information from LiDAR point clouds, while the GNSS/INS module provides a GNSS-aided inertial navigation solution as an additional reference. Finally, the EKF filter fuses these solutions, updates errors, and generates the final navigation solution. Furthermore, the feedback mechanism corrects IMU mechanization errors through a closed-loop compensation scheme, thereby enhancing the robustness and accuracy of the system.}
    \label{fig_LAE_GNSS_INS}
    \vspace{-3.5ex}
\end{figure}

\begin{table*}[!htbp]
    \centering
    \caption{Summary of the Collaboration of Positioning, Navigation, and Surveillance Technology}
    \label{table2}
    \begin{tabular}{c|c|c|c}
    \hline
    \textbf{} & \textbf{Technologies} & \textcolor{b}{\textbf{References}} & \textbf{Pros \& Cons} \\
    \hline
    \multirow{6}{*}{\shortstack{\textbf{High-Accuracy} \\ \textbf{Positioning}}}
    & \multirow{3}{*}{Improved GNSS} & \multirow{3}{*}{\cite{Xie2024, Tahir2019, Yao2022, Liu2024, Zhang2021}} & \checkmark Enhancing GNSS positioning accuracy with innovative auxiliary technologies \\
    & & & \checkmark Supporting precise analysis through simulation tool and generalized framework\\
    & & & \texttimes Failing to provide accurate positioning in satellite scarcity or signal blockage \\
    \cline{2-4}
    & \multirow{3}{*}{A-GNSS} & \multirow{3}{*}{\cite{van2020, AbuShaban2020, Li2023}} & \checkmark Accelerating signal acquisition and reducing TTFF  \\
    & & & \checkmark Minimizing signal propagation errors and improving positioning accuracy \\
    & & & \texttimes Relying on the availability and quality of terrestrial communication networks\\
    \hline
    \multirow{3}{*}{\shortstack{\textbf{Dynamic-Fusion} \\  \textbf{Navigation}}}
    & \multirow{3}{*}{Multi-Sensor} & \multirow{3}{*}{\cite{Wu2024, s22124327, Liu2021}} & \checkmark Overcoming the limitations of single sensors by combining different sensor strengths \\
    & & & \checkmark Improving localization accuracy and system robustness \\
    & & & \texttimes \shortstack{Facing higher complexity, sensor noise, and data sync issues} \\
    \hline
    \multirow{9}{*}{\shortstack{\textbf{Real-Time}\\ \textbf{Surveillance}}}
    & \multirow{3}{*}{\shortstack{Encryption \\ and \\ Authentication}} & \multirow{3}{*}{\cite{Wu2020, Ngamboe2025}} & \checkmark Reducing the risk of unauthorized message injection and illegal access \\
    & & & \checkmark Preventing tampering or falsification and improving system security.\\
    & & & \texttimes Resulting in additional computational overhead and bandwidth consumption \\
    \cline{2-4}
    & \multirow{3}{*}{Advanced \textcolor{b1}{TDoA}} & \multirow{3}{*}{\cite{Strohmeier2018, ELMARADY2021217}} & \checkmark Validating the positioning information in ADS-B messages\\
    & & & \checkmark Reducing dependence on time synchronization \\
    & & & \texttimes Suffering from lower positioning accuracy, especially in complex environments\\
    \cline{2-4}
    & \multirow{3}{*}{\shortstack{Intelligent Algorithms\\ and\\ Signal Processing}} & \multirow{3}{*}{\cite{Liu2024d, Fried2021}} & \checkmark  Strengthening the defense of the system against spoofing attacks \\
    & & & \checkmark Adapting well to dynamic and complex signal environments \\
    & & & \texttimes Requiring high data quality and computational resources \\
    \hline
    \end{tabular}
\end{table*}

\par 3) {\textit{\textbf{Real-Time Surveillance.}} Automatic dependent surveillance-broadcast (ADS-B) is a data broadcast-based air traffic surveillance technology, which is designed for real-time monitoring and collision avoidance of low-altitude aircraft \cite{DONG2024}. To be specific, ADS-B enables aircraft to periodically broadcast position, speed, heading, altitude, and other data via an onboard transmitter. This information is transmitted through radio frequencies and can be received by other aircraft and ground control stations, thereby allowing real-time tracking of the aircraft. Unlike traditional surveillance methods that rely on radar echoes \cite{Oh2021}, ADS-B provides precise location data based on the GNSS, with the ability to automatically broadcast, which offers advantages in terms of accuracy and timeliness. In addition, ADS-B features high precision, low latency, and wide coverage, so that it can offer reliable surveillance services over large areas. This makes it particularly suitable for complex low-altitude airspace, so that significantly improving the air traffic management efficiency and enhancing aircraft safety. Based on these advantages, \cite{Tong2023} developed a conflict awareness scheme for UAVs that analyzes the ADS-B message flow and format to identify potential conflicts. Next, strategies, such as speed adjustments and direction changes, based on the unscented Kalman filter were applied to resolve conflicts. This approach significantly improves the UAV conflict resolution capabilities, and provides a strong theoretical foundation for the advancement of collision avoidance technologies in UAVs.

\par However, the limitations of ADS-B should not be overlooked due to its openness and lack of built-in security mechanisms \cite{Leonardi2021}. Specifically, onboard GNSS receivers are vulnerable to interference or spoofing, and fake ADS-B messages can be injected or tampered with. Moreover, attackers can eavesdrop on broadcast messages or manipulate the ADS-B transponder of aircraft. These security vulnerabilities can lead to the distortion of surveillance data, thereby posing significant threats to air traffic safety. To mitigate the risks associated with ADS-B, several technologies have been proposed in recent years. By effectively authenticating and performing integrity checks on ADS-B messages, unauthorized tampering and message injection can be prevented, thereby protecting the system from attacks. For example, the author in \cite{Wu2020} presented an ADS-B message authentication approach based on certificate-free short signatures, which was designed to address security issues in ADS-B systems under low bandwidth and limited data bit conditions. Furthermore, \cite{Ngamboe2025} introduced the compatible authenticated bandwidth-efficient broadcast protocol for ADS-B (CABBA). However, this technology requires modifications to the ADS-B protocol, which would result in significant changes for existing aircraft and air traffic management systems.

\begin{figure}[!t]
    \centering
   \includegraphics[width=3.5in]{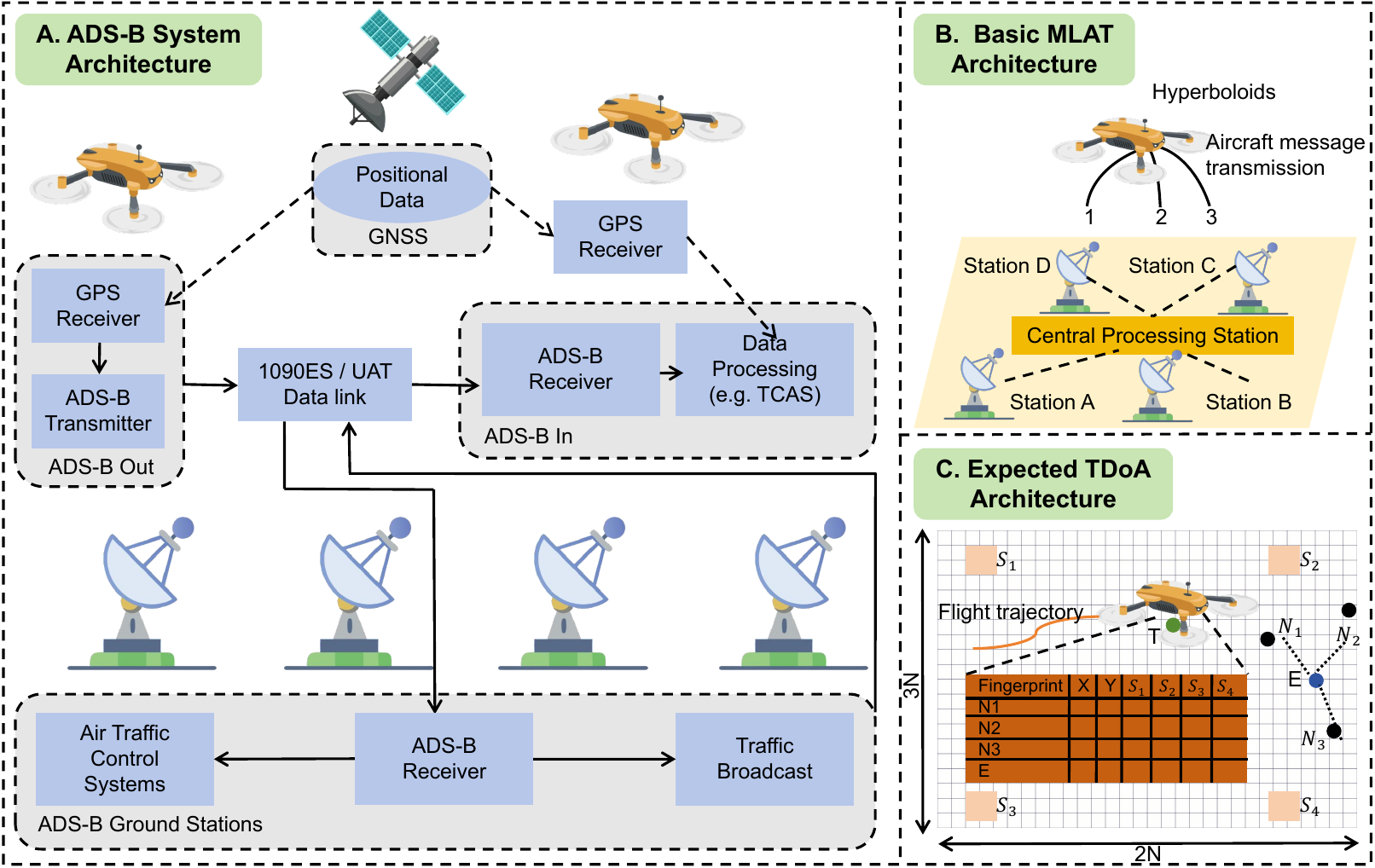}
    \caption{Air traffic surveillance architectures: ADS-B, MLAT, and expected \textcolor{b1}{TDoA}s \cite{Strohmeier2018}. \textit{Part A} describes the ADS-B system, which relies on GNSS to provide positional information and facilitates data exchange between aircraft and ground stations through ADS-B Out/In. \textit{Part B} introduces the MLAT architecture, which enhances surveillance capabilities by using multiple ground-based receivers to measure the \textcolor{b1}{TDoA} of signals and computes the actual position of the aircraft through a central processing station. \textit{Part C} presents the expected \textcolor{b1}{TDoA} architecture, which reduces the reliance on time synchronization by using \textcolor{b1}{TDoA} fingerprint maps with small grids.
}
    \label{fig_LAE_ADS-B}
    \vspace{-3.5ex}
\end{figure}

\par Additionally, relying solely on encryption and authentication approaches cannot fully address all security threats, particularly in the case of fake data injection. To mitigate this issue, researchers have proposed using multilateration (MLAT) \cite{Nijsure2016, Jheng2020} and time difference of arrival (\textcolor{b1}{TDoA}) \cite{Landzaat2024, Leonardi2019, Yang2021, Naganawa2021} techniques to validate the authenticity of position information in ADS-B messages. However, since MLAT suffers from lower accuracy in aircraft positioning, the expected \textcolor{b1}{TDoA} positioning technique has emerged to address this problem \cite{Strohmeier2018}. Specifically, this technique reduces the reliance on time synchronization by using \textcolor{b1}{TDoA} fingerprint maps with small grids (Fig. \ref{fig_LAE_ADS-B}). However, the positioning accuracy is still less than ideal, and it requires longer computation times. To overcome these limitations, an actual \textcolor{b1}{TDoA}-based augmentation system (ATBAS) was proposed \cite{ELMARADY2021217}, which uses data from the OpenSky network \cite{Strohmeier2018} to train \textcolor{b1}{TDoA} fingerprint grid models. Experimental results show that the ATBAS framework improves positioning accuracy by 56.93\% and 48.86\% over MLAT and the expected \textcolor{b1}{TDoA} technique, respectively. 

\par In response to the threats posed by GNSS spoofing and malicious ADS-B transmitters, researchers also have proposed several innovative protective approaches that combine intelligent algorithms with signal processing techniques to enhance system defense against these attacks. For instance, the author investigated a transfer learning-based approach to address the identification of malicious ADS-B transmitters \cite{Liu2024d}. In addition, the authors applied differential time-series transformation to process ADS-B data and employed non-cyclic autoencoders for attack classification and detection \cite{Fried2021}. This approach effectively identifies trajectory modification attacks, thereby improving the ability of the system to guard against GNSS spoofing. However, these techniques require high data quality and computational resources, and in certain complex signal scenarios or cases with significant data discrepancies, the accuracy and generalization ability of the model may be compromised.

\par As summarized in Table \ref{table2}, positioning, navigation, and surveillance each demonstrate significant effectiveness in high-precision positioning, dynamic navigation, and real-time surveillance, respectively. However, most existing research focuses primarily on positioning, navigation, surveillance, or combinations of two of these, without fully exploring the synergistic effects among all three. Thus, more research is needed to explore the deep integration of these technologies, thereby achieving higher precision, enhanced reliability, and greater autonomy in flight operations, particularly in the LAE networks.

\subsection{Fusion of Flight Control with Airspace Management}
\label{sec_Fusion}

\par This fusion is crucial for improving the efficiency and safety of low-altitude airspace operations. Flight control optimizes flight paths through intelligent scheduling to ensure smooth operations in high-density airspace, while airspace management optimizes real-time resource allocation to prevent conflicts and congestion. This integration of these technologies not only improves the operational efficiency, but also maximizes the use of airspace resources, thereby ensuring seamless coordination across multiple tasks.

\begin{figure}[!t]
    \centering
   \includegraphics[width=3.5in]{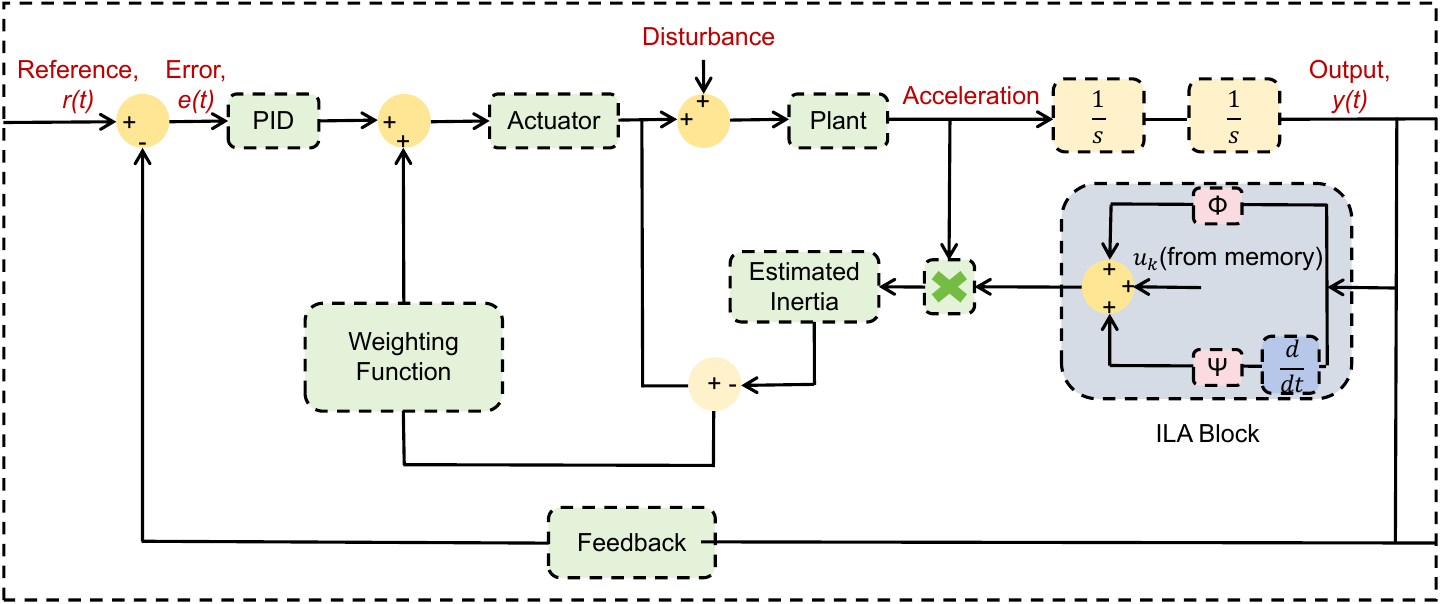}
    \caption{A schematic diagram of PID–ILAFC \cite{Abdelmaksoud2020}. Specifically, the PID controller generates control signals, while the ILA module compensates by storing and learning error information to improve the disturbance rejection and dynamic performance of the system. Moreover, the feedback loop corrects errors to ensure that the output $y(t)$ accurately tracks the reference input $r(t)$.}
    \label{fig_LAE_PID}
    \vspace{-3.5ex}
\end{figure}
\par 1) {\textit{\textbf{Advanced-Flight Control.}} Proportional integral differential (PID) control \cite{LopezSanchez2023} is one of the most common traditional flight control approaches. This method maintains aircraft stability by adjusting control inputs based on real-time feedback of attitude and position \cite{math11153390}. However, traditional PID control often struggles to address the challenges posed by nonlinear systems, external disturbances, and parameter tuning when faced with complex and dynamic flight tasks, thus limiting its performance. Therefore, adopting AI-driven PID control becomes crucial, as it enhances the adaptability and robustness of the system in complex environments. For example, \cite{mohammed2023} proposed an approach that improves the traditional PID controller by using a feedforward neural network. This approach not only maintains the stability of the quadrotor but also overcomes some of the hardware implementation limitations of traditional PID control, so that improving the accuracy and performance of the flight control system. However, as AI advances in aircraft control, using only PID control is inadequate for the complex flight tasks, especially with external disturbances and internal uncertainties.

\par To address these challenges, a robust intelligent self-adjusting propulsion control approach has emerged. For instance, the authors in \cite{Abdelmaksoud2020} developed a hybrid control approach combining PID and intelligent active force control (IAFC) to enhance the disturbance rejection capability and agility performance of a quadrotor UAV. To further enhance robustness, an AI-based iterative learning algorithm (ILAFC) was used to automatically adjust control parameters (Fig. \ref{fig_LAE_PID}), which results in a 17\% improvement in disturbance rejection and agility performance. In addition to optimizing the control strategy, the computational bottleneck faced by the aircraft during target detection and tracking tasks also needs to be addressed. For example, \cite{Rabah2020a} introduced a heterogeneous parallelization strategy that integrates target detection and tracking algorithms with UAV control software. Then, based on target tracking data, a gain-scheduled PID controller further optimizes flight control to ensure efficient task execution in complex environments. Therefore, AI-based PID control approaches significantly improve the UAV performance in certain tasks such as disturbance rejection, target tracking, and agility. Similarly, in LAE applications, AI-based control strategies play a key role in optimizing UAV operational efficiency, energy consumption, and the overall system robustness.

\begin{table*}[!htbp]
    \centering
    \caption{Summary of the Fusion of Flight Control with Airspace Management Technology}
    \label{table3}
    \begin{tabular}{c|c|c|c}
    \hline
    \textbf{} & \textbf{Technologies} & \textcolor{b}{\textbf{References}} & \textbf{Pros \& Cons} \\
    \hline
    \multirow{6}{*}{\shortstack{\textbf{Advanced-Flight} \\ \textbf{Control}}}
    & \multirow{3}{*}{AI-based PID} & \multirow{3}{*}{\cite{mohammed2023, Abdelmaksoud2020, Rabah2020a}} & \checkmark Effectively addressing external disturbances and internal uncertainties\\
    & & & \checkmark Improving the accuracy and performance of the flight control system\\
    & & & \texttimes Struggling to maintain stability in rapidly changing and uncertain environments\\
    \cline{2-4}
    & \multirow{3}{*}{AI-based MPC} & \multirow{3}{*}{\cite{van2020, AbuShaban2020, Li2023}} & \checkmark Considering future states and constraints to improve decision-making \\
    & & & \checkmark Enabling flexible and adaptive control in complex and dynamic environments \\
    & & & \texttimes Involving high computational demands for real-time optimization and state prediction\\
    \hline
    \multirow{8}{*}{\shortstack{\textbf{Efficient-Airspace} \\  \textbf{Management}}}
    & \multirow{2}{*}{\shortstack{Fully Mixed\\Management}} & \multirow{2}{*}{\cite{sunil2016}} & \checkmark Relying on global coordination and real-time information to ensure safe separation\\
    & & & \texttimes Increasing coordination difficulty in high-demand environments \\
    \cline{2-4}
    & \multirow{2}{*}{\shortstack{Layered\\Management}} & \multirow{2}{*}{\cite{sunil2016, Sarim2019, McCarthy2020}} & \checkmark Reducing aircraft conflicts by dividing airspace into layers\\
    & & & \texttimes Unevenly distributing airspace may reduce capacity and affect noise control efficiency \\
    \cline{2-4}
    & \multirow{2}{*}{\shortstack{Zoning\\Management}} & \multirow{2}{*}{\cite{Gharibi2020}} & \checkmark Enabling different flight tasks to proceed simultaneously\\
    & & & \texttimes Requiring complex coordination for inter-regional flights, and increasing noise levels \\
    \cline{2-4}
    & \multirow{2}{*}{\shortstack{Corridor\\Management}} & \multirow{2}{*}{\cite{Low2016, Pathiyil2016}} & \checkmark Minimizing conflicts and noise through fixed flight paths\\
    & & & \texttimes Lacking flexibility, which may hinder adaptability in complex flight environments\\
    \hline
    \end{tabular}
\end{table*}

\par Model predictive control (MPC) is another advanced flight control approach that predicts the future state of an aircraft and generates control inputs through optimization to make optimal control decisions \cite{kouvaritakis2015}. Unlike the PID control, which focuses on current errors, MPC also considers changes in future states and constraints. For instance, the authors in \cite{2023constrained} proposed a general-model dynamic formulation and a two-layered constrained MPC strategy to tackle the trajectory tracking problem for tilt-rotor UAVs. Moreover, a novel nonlinear model predictive control (NMPC) approach has been designed to address UAV navigation and obstacle avoidance \cite{Lindqvist2020}. While these approaches perform well in trajectory tracking and obstacle avoidance tasks, integrating AI-based MPC can further enhance their performance, particularly in improving the accuracy of state predictions and optimizing control decisions.

\begin{figure}[!t]
    \centering
   \includegraphics[width=3.5in]{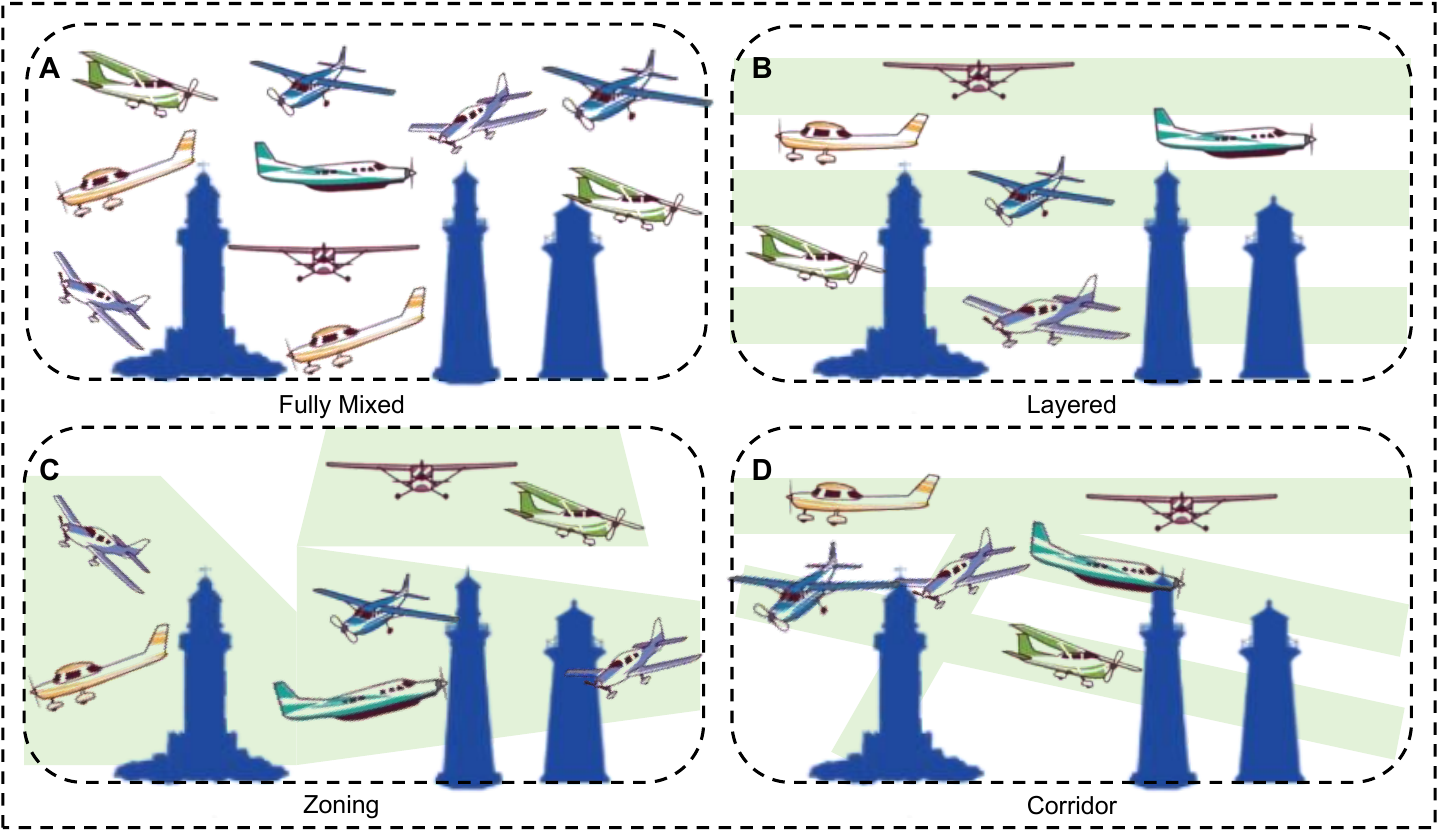}
    \caption{\textcolor{b}{An illustration of airspace management for low-altitude aircraft. \textit{Part A} shows the fully mixed management, \textit{Part B} presents the layered management, \textit{Part C} depicts the zoning management, and \textit{Part D} demonstrates the corridor management.}}
    \label{fig_LAE_ATM}
    \vspace{-3.5ex}
\end{figure}

\par AI can optimize the aircraft model by learning from the historical data, such that improving the accuracy of state predictions and adjusting control decisions in real-time, which enables the aircraft to respond more flexibly and quickly to environmental changes. For example, \cite{Zhang2024} proposed a UAV control approach that combines neural network learning with NMPC to enhance the tracking accuracy of the aircraft under external disturbances and internal structural changes. Moreover, the authors introduced a novel framework integrating MPC and DRL methods, where the MPC component operates at a lower frequency to provide baseline control inputs, while the DRL component adjusts the outputs generated by MPC at a higher frequency \cite{Sun2024b}. These \textcolor{b1}{highlight} that the AI-based MPC has become a critical tool for enhancing aircraft performance, and it is an essential technology in LAE networks. By learning and adapting in real time, AI-based MPC can ensure that the aircraft performs stable and efficient tasks under varying factors such as wind changes, complex terrain, and dynamic obstacles. This makes it suitable for a wide range of applications, including urban air mobility, aerial logistics, agricultural monitoring, and environmental surveillance.

\par 2) {\textit{\textbf{Efficient Airspace Management.}}} Air traffic management (ATM) focuses on the scheduling and management of the low-altitude airspace to ensure the safe operation of aircraft and the efficient utilization of airspace resources \cite{Garrow2021}. Several airspace management models have been proposed and studied, \textcolor{b}{as shown in Fig. \ref{fig_LAE_ATM}}, each with its unique advantages and challenges in terms of flight safety and airspace efficiency. First, the fully mixed management model is an unstructured approach, where the aircraft do not follow the fixed flight paths \cite{sunil2016}. Instead, it relies on global coordination and real-time information sharing to maintain safe separation between aircraft. While this model provides flexibility and accommodates various flight demands, the lack of a clear structural framework may introduce complexity in management, thereby reducing the efficiency, particularly in high-demand scenarios where the coordination becomes a limiting factor. \textcolor{b}{In such cases, the model can benefit from real-time data exchange supported by 5G-A networks, dedicated air-to-air communication channels, satellite communication, and ground relay systems, which help resolve conflicts dynamically and enhance overall efficiency.} Second, layered management divides airspace into multiple altitude layers to reduce conflicts between aircraft. For example, \cite{sunil2016} suggested that each altitude layer can be assigned specific flight speeds to improve flight efficiency. The authors in \cite{Sarim2019} divided the airspace below 400 feet into several layers based on the minimum safety operational diameter. Moreover, a three-layer airspace design was proposed, where each layer group includes both conflict-free and operational layers \cite{McCarthy2020}. However, layered management may be affected by the uneven distribution of airspace resources, which could reduce the overall capacity and lead to relatively poor noise control. \textcolor{b}{To enhance this model, real-time monitoring and dynamic adjustments across layers can be implemented. For instance, cloud-based air traffic management systems can assess traffic conditions as they occur and re-assign aircraft between layers to alleviate congestion and optimize resource allocation.}

\begin{figure}[!t]
    \centering
   \includegraphics[width=3.5in]{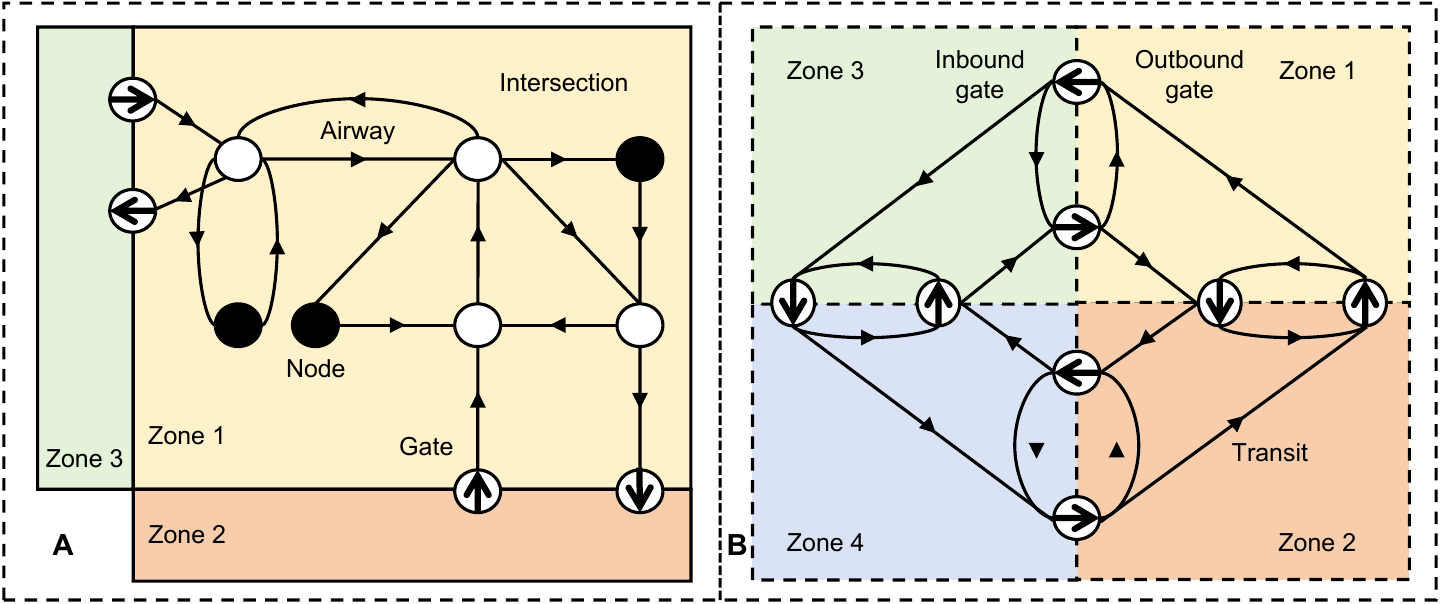}
    \caption{The network structure based on zones \cite{Gharibi2020}. \textit{Part A} focuses on the structure and flow within a single zone, thereby illustrating how movement occurs through nodes and gates within the zone. \textit{Part B} shows the overall structure across multiple zones, where gates serve as entry and exit points between different zones. Moreover, transit edges represent the movement of UAVs from an inbound gate in one zone to an outbound gate in another.}
    \label{fig_LAE_Zone}
    \vspace{-3.5ex}
\end{figure}

\begin{figure*}[!hbt] 
	\centering
	\includegraphics[width =7in]{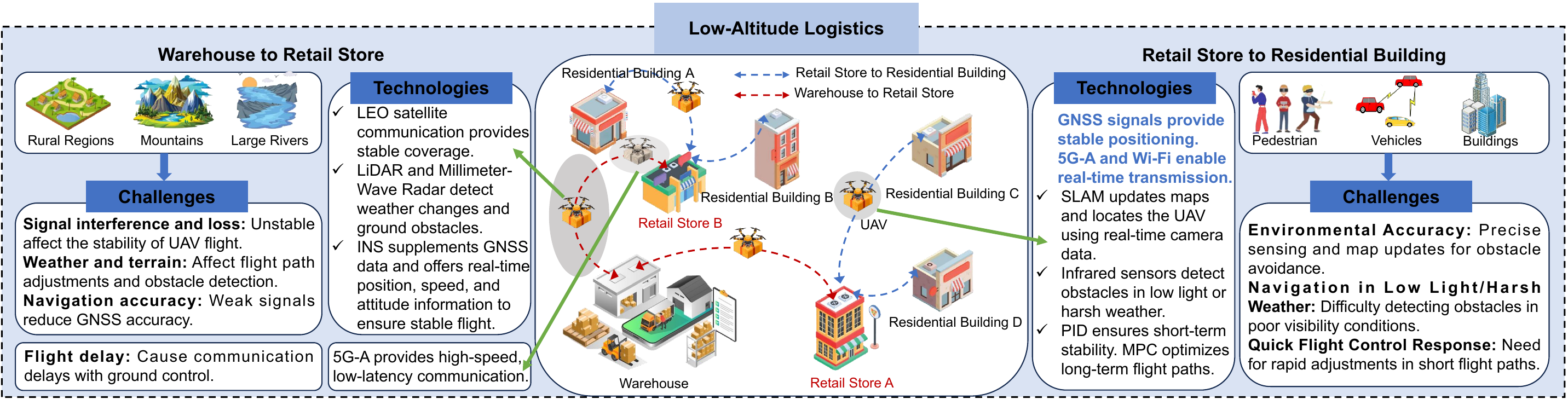}
	\caption{The role of multi-technology integration in optimizing UAV operations for logistics, which focuses on two key scenarios: warehouse to retail store and retail store to residential building.}
	\label{fig_LAE_Logistics}
    \vspace{-1.2em}
\end{figure*}

\par Additionally, the zoning management model divides airspace into multiple independent regions, allowing different flight operations to occur simultaneously. For instance, in \cite{Gharibi2020}, the authors partitioned the airspace into several regions, with each region managed by a zone service provider (ZSP). Within each region, individual spaces offer navigation information and flight instructions to ensure safety (Fig. \ref{fig_LAE_Zone}). Although this model helps optimize capacity, safety, and efficiency, flying across regions requires complex coordination and may result in higher noise levels. \textcolor{b}{In this context, seamless communication between ZSPs is essential and can be achieved through automated communication protocols (e.g., API-driven service meshes) and edge computing platforms, while AI-enhanced decision-making frameworks streamline cross-zone coordination for uninterrupted transitions.} Finally, the corridor management approach minimizes conflicts and noise in the airspace by providing fixed flight paths for aircraft. The authors designed an airspace system with multiple bidirectional traffic corridors, each separated by a safe distance \cite{Low2016}. Furthermore, the authors introduced a tunnel airspace concept specifically tailored for UAVs, which enables dynamic design and regulates UAV traffic through no-fly zones \cite{Pathiyil2016}. However, the drawback of this management model is its lack of flexibility, which limits the adaptability for dynamic flight path adjustments. \textcolor{b}{To overcome this issue, integrating ground-to-air (G2A) communication networks and multi-constellation GNSS navigation systems can enable adaptive corridor reconfiguration based on real-time demand, thus maintaining safety margins during peak operations.} In conclusion, different airspace management models have varying applicability depending on the scenario and specific needs. Choosing the appropriate model can significantly enhance the efficiency and safety of low-altitude airspace utilization, which is crucial for the development of the LAE networks.

\par In summary, the fusion of flight control and airspace management plays a critical role in improving the safety and efficiency of low-altitude airspace operations. While the key technologies required for each have been outlined in Table \ref{table3}, existing research has yet to explore the comprehensive fusion of these two areas in depth. Future research should prioritize their joint implementation to enhance flight accuracy, optimize resource distribution, and improve system flexibility, particularly within the scope of the LAE networks, where the effective integration is essential for seamless operations and the optimal use of airspace.

\section{Multi-Technology Integration for Low-Altitude Logistics, Traffic, and Rescue}
\label{sec_application}

\par In this section, we explore various applications of multi-technology integration in real-world scenarios, such as logistics (Fig. \ref{fig_LAE_Logistics}), rescue (Fig. \ref{fig_LAE_Rescue}), and transportation (Fig. \ref{fig_LAE_transportation}).
%as presented in Figs. \ref{fig_LAE_Lofistics}.

% \begin{figure*}[!hbt] 
% 	\centering
% 	\includegraphics[width =6.5in]{figure//LAE_app.pdf}
% 	\caption{The role of multi-technology integration in optimizing UAV operations for logistics, rescue, and transportation. By integrating these technologies, UAVs can enhance efficiency, safety, and adaptability across various operational environments.}
% 	\label{fig_LAE_app}
%     \vspace{-1.2em}
% \end{figure*}

\subsection{Low-Altitude Logistics}
\par When UAVs perform mid to long distance transportation (such as from a warehouse to a retail store), the goal is to efficiently and stably transport goods across regions and long distances. Specifically, during the remote stages of the flight, UAVs may pass through the areas with poor signal coverage, such as rural regions, mountains, or large rivers. In these cases, 5G-A may struggle with signal interference or loss, while LEO satellite communication can maintain consistent coverage \cite{LeyvaMayorga2020}. As the UAVs approach the retail store in an urban area, 5G-A offers high-speed and low-latency communication, which allows the UAVs to quickly respond to ground commands and transmit flight status data in real time \cite{Masaracchia2021}. Moreover, radar systems detect obstacles at greater distances and accurately map the flight path to enable dynamic obstacle avoidance. For example, both LiDAR and millimeter-wave radar detect the height and distance of ground obstacles and sense weather changes (such as cloud cover and wind speed) \cite{RAMASAMY2016, Soumya2023}, thereby enabling the UAVs to adjust their flight trajectory in real time to avoid collisions with these natural obstacles. Furthermore, to ensure precise navigation during long-range flights, GNSS provides support so that UAVs can navigate accurately in areas with ground station signal coverage. However, when flying through the areas with signal obstructions or unstable conditions, INS \cite{Gyagenda2022} continuously provides real-time data on the position, velocity, and attitude of the UAVs via IMUs.

\begin{figure}[!t] 
	\centering
	\includegraphics[width =3.5in]{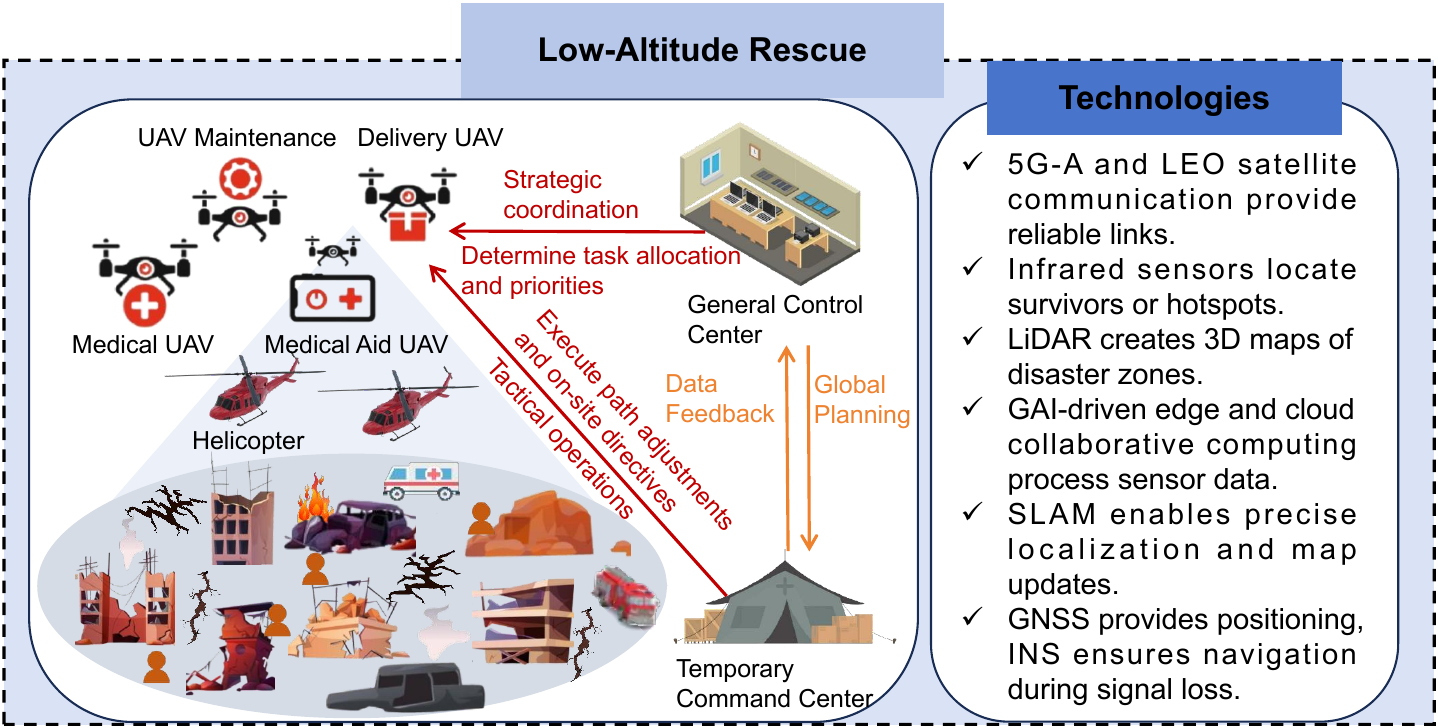}
	\caption{Leveraging multi-technology integration to enhance UAV efficiency in rescue operations. Advanced technologies like 5G-A, LiDAR, SLAM, and GAI-driven computing enable real-time situational awareness, precise localization, and efficient task coordination.}
	\label{fig_LAE_Rescue}
    \vspace{-3.5ex}
\end{figure}

\par In short-distance deliveries (such as from a retail store to a residential building), the goal is to quickly and efficiently complete the ``last-mile" \cite{Borghetti2022}, which ensures that the UAVs reach their destination precisely and safely within a small area. In particular, GNSS signals are typically stable enough to provide global positioning support, and 5G-A or Wi-Fi networks are sufficient to support real-time data transmission and command updates. The focus shifts to more precise environmental awareness and flight control. For instance, in urban or residential areas, UAVs face numerous dynamic obstacles, such as pedestrians, vehicles, and buildings \cite{Shakhatreh2019}. To avoid collisions, SLAM captures real-time environmental data with cameras, thereby updating the map and accurately determining the position of UAVs \cite{Gupta2022}. Infrared sensors detect heat sources and identify invisible obstacles, thus improving navigation in low-light or adverse weather conditions \cite{Nguyen2021}. In addition, since the path is short, the flight control system must respond quickly. For example, when a UAV approaches an obstacle (e.g., a pedestrian or parked vehicle), the system calculates an avoidance path and adjusts the trajectory, such as changing altitude or shifting laterally, to maintain a safe distance. The system also considers weather and wind conditions to optimize the flight path in real time.

\subsection{Low-Altitude Rescue}
\par The key to this application lies in rapid response, precise positioning, and environmental awareness, which guarantees the UAVs can quickly reach target areas and execute tasks in emergency situations. To be specific, 5G-A and LEO satellite communications provide reliable communication links between UAVs and rescue command centers \cite{Wang2023a}. This ensures that the UAVs can receive instructions and transmit data continuously in complex or remote rescue scenarios. This is crucial for real-time tracking of rescue progress and sharing on-site information, such as video streams and environmental data. Moreover, infrared sensors and LiDAR offer precise environmental perception in low-light or harsh weather conditions, thereby enabling the detection of potential obstacles, assessing disaster-affected areas, and identifying targets in the need of rescue. For example, infrared sensors detect heat sources \cite{Shi2024}, such as the body temperature of trapped individuals, to locate survivors or hot spots, so that the UAVs can identify targets in low-light or poor-visibility conditions. In addition, LiDAR provides 3D mapping of disaster zones to help rescue teams plan operations \cite{Lee2016}. 

\par The information gathered by these sensors is then processed by GAI-driven edge computing and cloud collaborative computing. More precisely, edge computing quickly analyzes data on the UAVs or nearby devices, which allows real-time adjustments to flight paths, obstacle avoidance, and target localization \cite{Demiane2020}. Cloud collaborative computing integrates large-scale data and shares it with other devices to generate global disaster maps, optimize task allocation, and develop overall rescue strategies, such that significantly improving the rescue efficiency \cite{Shah2023, Sun2024c}. Furthermore, SLAM enables the UAVs to perform precise localization and map updates even in the areas lacking stable GPS signals, such as underground ruins or deep ravines after an earthquake \cite{Schleich2021}. To guarantee stable and accurate flight paths, GNSS provides broad global positioning support for UAVs in open areas, while INS supplements GNSS when signals weaken or are lost, thereby ensuring continuous navigation for short periods.

\begin{figure}[!t] 
	\centering
	\includegraphics[width =3.5in]{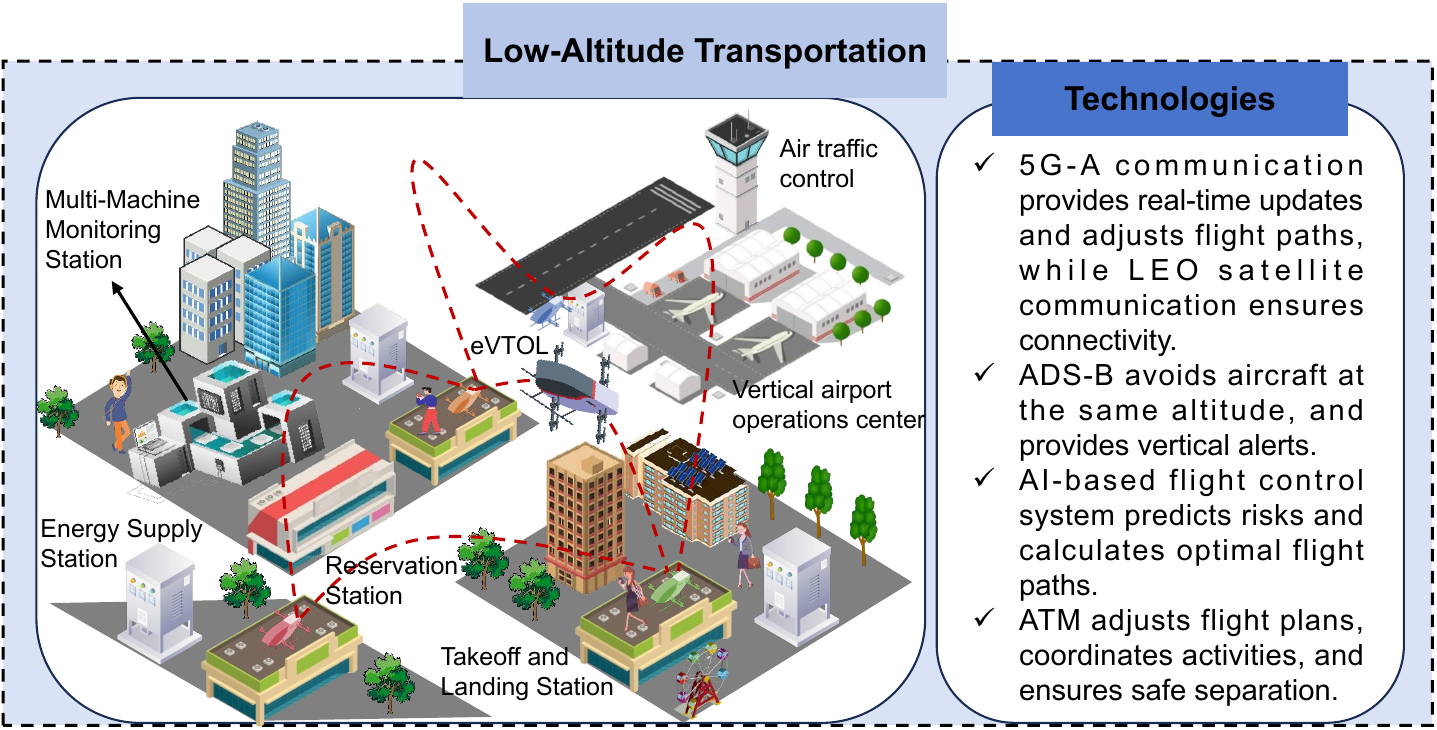}
	\caption{Enhancing UAV transportation operations through multi-technology integration. By leveraging 5G-A, LEO satellite communication, ADS-B, and AI-based flight control, UAVs achieve real-time path optimization, air traffic coordination, and safe low-altitude navigation.}
	\label{fig_LAE_transportation}
    \vspace{-1.2em}
\end{figure}

\subsection{Low-Altitude Transportation}
\par In this application, ensuring safe interaction between aircraft and airspace coordination is crucial. With its low latency and high bandwidth, 5G-A communication enables aircraft to receive real-time updates on airspace changes, weather conditions, and flight plans, which allows for prompt adjustments to their flight paths. Moreover, LEO satellite communication ensures that the aircraft maintain connectivity with ground control centers, even in remote or signal-restricted areas. At the same time, GAI-driven cloud collaborative computing supports the storage and processing of large-scale real-time data, such as traffic flow, aircraft positions, and airspace conditions \cite{Huang2024a}, thereby facilitating real-time traffic monitoring, intelligent decision support, and various computational services. 

\par Moreover, in congested airspace, ADS-B technology helps aircraft detect and avoid other aircraft at the same altitude, while also providing vertical alerts to prevent potential conflicts. For example, when the trajectories of two aircraft are about to intersect, the ADS-B provides real-time data on their positions and speeds, thereby enabling aircraft to adjust paths and avoid possible collisions \cite{wu2023}. Therefore, ADS-B compensates for the limitations of visual perception and enhances safety for aircraft. Furthermore, AI-based flight control systems use real-time sensor data and airspace information to predict potential risks and calculate the optimal flight path based on the dynamics of aircraft. In addition, ATM dynamically adjusts flight plans based on the position, speed, and task requirements of aircraft, which coordinates activities between different aircraft to ensure efficient use of airspace resources and safe separation \cite{GARDI2016}. ATM also provides real-time tracking and scheduling to ensure smooth navigation through complex airspace environments.

{\color{b}
\subsection{Lessons Learned}
\label{sec_Lessons Learned_}
\par Exploring the applications of multi-technology integration in low-altitude logistics, rescue, and transportation offers valuable insights. Specifically, in logistics, the integration of multiple technologies significantly enhances UAV efficiency and safety for both long-distance and short-distance deliveries. In rescue scenarios, the synergy of technologies strengthens UAV positioning and environmental awareness under challenging conditions, thereby improving real-time decision-making and task coordination. Moreover, in transportation, the combined technologies ensure aircraft safety and coordination in crowded airspace, while ATM maximizes the utilization of airspace resources. These cases demonstrate that complementary technologies provide significant improvements in efficiency, safety, and adaptability compared to single technologies. However, overcoming complex integration challenges, high costs, and reliability issues is crucial to unlocking their full potential in diverse applications.
}

\section{Future Directions}
\label{sec_Future Directions}
\par In this section, we will present \textcolor{b}{important} future directions for collaborative technologies in LAE networks.

\subsection{Intelligent and Adaptive Optimization of Dynamic Airspace}
\par The coexistence of high-density, multi-type aircraft in low-altitude airspace requires overcoming traditional air traffic control constraints by establishing an intelligent and adaptive management system. By leveraging GAI-enabled DRL or game theory within a distributed decision-making framework and utilizing digital twin platforms \cite{Lehner2022} alongside various low-altitude sensors and communication devices, the system can monitor aircraft status in real time and create an intelligent traffic model that dynamically adjusts flight routes, speed, or altitude to avoid congestion. Moreover, airspace usage rights can be managed through an auction model \cite{Leet2024} that incorporates different time slots and priority tasks to design differentiated pricing strategies. Moreover, blockchain smart contracts \cite{Wang2019} can then be used to automatically execute transactions, thus incentivizing users to actively avoid peak airspace and optimizing flight management within the LAE networks.

\subsection{Security and Privacy Protection of LAE Networks}
\par The openness and data mobility of LAE networks expose them to multiple security threats, which necessitate a multi-layered defense system. For example, the cloud-edge-end collaborative security architecture can protect data confidentiality and integrity by deploying trusted execution environments (TEEs) on aircraft \cite{Valero2022}, running intrusion detection systems (IDS) at edge nodes \cite{Eskandari2020}, and utilizing homomorphic encryption at the cloud level \cite{Mahato2021}. In terms of the privacy protection, LAE networks involve substantial amounts of user data (e.g., flight trajectories and personal identities). Therefore, techniques such as differential privacy \cite{Wang2022} and federated learning \cite{choudhury2020} can anonymize and protect user data, thus ensuring that the privacy is not compromised during data sharing and analysis. \textcolor{b2}{To further enhance the security and privacy, network resilience is required to handle unexpected events such as cyberattacks or equipment failures. Thus, AI-based anomaly detection \cite{DeMedeiros2023}, autonomous reconfiguration mechanisms \cite{Bradley2020}, and multi-path communication technologies \cite{Liu2024t} can be employed to improve system adaptability and recovery capabilities, so that continuous protection of critical data and services is ensured.}

\subsection{Sustainable Power and Energy Management of Aircraft}
\par The development of LAE networks may face challenges in sustainable energy supply, particularly with the large-scale deployment of high-energy-consuming aircraft such as eVTOLs and UAVs, which require the enhanced dynamic response and sustainability in energy systems. Future research can focus on aerial energy relay networks \cite{Padilla2020}, where solar-powered UAVs function as the charging stations to create an air-to-air energy supply chain. Furthermore, bio-hybrid energy systems could be explored \cite{Fukui2025}, which use microbial fuel cells along with rainwater and carbon dioxide collected from aircraft surfaces to generate power. MmWave beamforming and laser energy transmission \cite{Jin2019} could enable wireless charging of aircraft during hovering or low-speed cruising. In addition, energy consumption prediction models that couple flight tasks, weather conditions, and battery health status could be developed, with GAI-enabled DRL used to optimize power distribution.

{\color{b}
\subsection{Quantum-Driven Intelligence Coordination for Airspace}
\par As LAE networks evolve toward massive-scale aircraft coordination in real time, classical computational paradigms may struggle to handle the exponential growth of state-action spaces and multi-agent constraints. To address this, integrating quantum computing, particularly quantum annealing \cite{rajak2023} and variational quantum circuits \cite{gong2024}, into airspace management makes it possible to accelerate route optimization, task allocation, and conflict resolution in high-density low-altitude environments. In particular, quantum-inspired algorithms can be embedded into digital twin platforms to simulate thousands of parallel flight scenarios under uncertainty \cite{gharehchopogh2023}. Moreover, hybrid classical-quantum models can support the optimization of secure and ultra-fast communication protocols by leveraging quantum entanglement properties \cite{cavaliere2020}, which in turn strengthens the resilience of collaborative decision-making among aircraft clusters in contested or congested airspace.

\subsection{Self-Evolving Cognitive Agents for Generative Governance}
\par Future low-altitude airspace systems will benefit from the deployment of self-evolving cognitive agents that continuously learn, adapt, and optimize their behavior across communication, control, and navigation layers. By integrating multi-modal data perception with generative behavior modeling \cite{chen2023b}, each agent can construct its own decision heuristics, which allows it to develop new coordination strategies based on collective experience \cite{chen2023c}. These agents are first trained within large-scale airspace simulations and are then progressively adapted through real-time in situ learning, so that they can operate effectively in dynamic environments. As a result, such generative governance frameworks \cite{taeihagh2025} can autonomously handle unknown traffic conditions, unexpected tasks, or policy changes. This capability helps to establish a decentralized yet coherent airspace control paradigm that minimizes human intervention while improving operational safety, efficiency, and fairness.
}

{\color{b2}\subsection{LAE-LEO Collaboration for 3D Airspace Coverage}
\par As low-altitude airspace tasks grow and coverage expands, the integration of LAE networks with LEO satellite networks will be essential for achieving seamless 3D coverage. Specifically, the dynamic orbits, low latency, and wide coverage of LEO satellites facilitate reliable cross-regional communication for LAE systems, particularly in areas with limited ground infrastructure or under extreme environmental conditions \cite{Ibraheem2024}. To support dynamic cross-domain collaboration, future research could focus on developing an intelligent and cooperative spectrum management architecture. This architecture integrates AI-based spectrum prediction \cite{Ramesh2021} and interference modeling \cite{Zhang2023a} to identify heterogeneous and multi-source interference in real time, so that spectrum reconfiguration and channel scheduling can be conducted based on the status of uplink and downlink channels. Furthermore, a unified air-ground-space digital twin system can be established to realize resource virtualization and link optimization between aircraft and LEO satellites \cite{He2023}. Such a system will facilitate the construction of a highly reliable and reconfigurable 3D collaborative network, thereby enabling task-level self-organization and intelligent interconnection in next-generation airspace operations.
}
}
\section{Conclusion}
\label{sec_Conclusion}
\par In this article, we conducted a comprehensive analysis of specific aspects of LAE networks. We began by describing the distinct advantages of LAE networks over traditional UAV networks and introduced the relevant standards and core architecture that support their development. Next, we investigated the integration of key technologies within LAE networks, including the integration of communication, sensing, and intelligent computing, collaboration of positioning, navigation, and surveillance, and the fusion of flight control with airspace management. Following this, we demonstrated that how these integrated technologies enable the sustainable operation of real-world LAE applications. Finally, we outlined the future research directions and highlighted the potential of collaborative technologies in LAE networks for intelligent and adaptive optimization, security and privacy protection, sustainable energy and power management, \textcolor{b2}{quantum-driven coordination, generative governance, and 3D airspace coverage}.

\ifCLASSOPTIONcaptionsoff
\newpage
\fi

\bibliographystyle{IEEEtran}
\bibliography{references.bib}

\end{document}